

\let\AMSfonts1 

\magnification 1200
\baselineskip=12pt
\hsize=15.3truecm \vsize=22 truecm \hoffset=.1truecm
\parskip=14 pt  \overfullrule=0pt
\def\nl{\hfill\break}

\def\titem#1\par{\item{$\triangleright$} #1\par \vskip-8pt}

\def\idty{{\leavevmode{\rm 1\mkern -5.4mu I}}}
\def\Ibb #1{ {\rm I\mkern -3.6mu#1}}
\def\Ird{{\hbox{\kern2pt\vbox{\hrule height0pt depth.4pt width5.7pt
    \hbox{\kern-1pt\sevensy\char"36\kern2pt\char"36} \vskip-.2pt
    \hrule height.4pt depth0pt width6pt}}}}
\def\Irs{{\hbox{\kern2pt\vbox{\hrule height0pt depth.34pt width5pt
       \hbox{\kern-1pt\fivesy\char"36\kern1.6pt\char"36} \vskip -.1pt
       \hrule height .34 pt depth 0pt width 5.1 pt}}}}
\def\Ir{{\mathchoice{\Ird}{\Ird}{\Irs}{\Irs} }}
\def\ibbt #1{\leavevmode\hbox{\kern.3em\vrule
     height 1.5ex depth -.1ex width .2pt\kern-.3em\rm#1}}
\def\ibbs#1{\hbox{\kern.25em\vrule
     height 1ex depth -.1ex width .2pt
                   \kern-.25em$\scriptstyle\rm#1$}}
\def\ibbss#1{\hbox{\kern.22em\vrule
     height .7ex depth -.1ex width .2pt
                   \kern-.22em$\scriptscriptstyle\rm#1$}}
\def\ibb#1{{\mathchoice{\ibbt #1}{\ibbt #1}{\ibbs #1}{\ibbss #1}}}
\def\Nl{{\Ibb N}} \def\Cx {{\ibb C}} 
\def\lessblank{\parskip=5pt \abovedisplayskip=2pt
          \belowdisplayskip=2pt }
\def\eproclaim{\par\endgroup\vskip0pt plus100pt\noindent}
\def\proof#1{\par\noindent {\bf Proof #1}\          
         \begingroup\lessblank\parindent=0pt}
\def\QED {\hfill\endgroup\break
     \line{\hfill{\vrule height 1.8ex width 1.8ex }\quad}
      \vskip 0pt plus100pt}
\def\Aut{{\rm Aut}}
\def\Bar{\overline}
\def\abs #1{{\left\vert#1\right\vert}}
\def\ad{{\rm ad}}
\def\bra #1>{\langle #1\rangle}
\def\bracks #1{\lbrack #1\rbrack}
\def\dim {\mathop{\rm dim}\nolimits}
\def\id{\mathop{\rm id}\nolimits}
\def\half{{\scriptstyle{1\over2}}}

\def\ket #1>{\mid#1\rangle}
\def\midbox#1{\qquad\hbox{#1}\qquad}  
\def\norm #1{\left\Vert #1\right\Vert}
\def\order{{\bf o}}
\def\rstr{\hbox{$\vert\mkern-4.8mu\hbox{\rm\`{}}\mkern-3mu$}}
\def\set #1{\left\lbrace#1\right\rbrace}

\def\stt{\,\vrule\ }
\def\th{\hbox{${}^{{\rm th}}$}\ }  
\def\tr{\mathop{\rm tr}\nolimits}
\def\teedis{\hbox{\kern1pt$\bigcirc$ \kern-12.3pt
     \lower1.5pt\hbox{$\top$} }}
\def\teescr{\hbox{\kern.5pt$\scriptstyle\bigcirc$ \kern-10.65pt
     \lower1.5pt\hbox{$\scriptstyle\top$} \kern-1pt}}
\def\tee{{\mathchoice{\teedis}{\teedis}{\teescr}{\teescr}}}
\def\eetdis{\hbox{\kern1pt$\bigcirc$ \kern-12.3pt $\perp$ }}
\def\eetscr{\hbox{\kern.5pt$\scriptstyle\bigcirc$ \kern-10.65pt
     $\scriptstyle\perp$ \kern-1pt}}
\def\eet{{\mathchoice{\eetdis}{\eetdis}{\eetscr}{\eetscr}}}
\def\biggerotimes{\hbox{\kern1pt$\bigcirc$ \kern-12.3pt $\times$ }}
\def\ot{{\mathchoice{\biggerotimes}{\biggerotimes}
        {\otimes}{\otimes}}}
\def\phi{\varphi}            
\def\epsilon{\varepsilon}    
\def\A{{\cal A}} \def\B{{\cal B}} \def\C{{\cal C}} \def\D{{\cal D}}
\def\G{{\cal G}} \def\H{{\cal H}} \def\K{{\cal K}} 
\def\M{{\cal M}}  
   
\def\E{{\Ibb E}}  \def\Eh{{\hat{\Ibb E}}} 
 
\def\cfc{C*-finitely correlated}
\def\cp{completely positive}
\def\pg{purely generated}
\def\ie{i.e.\ }               
\def\eg{e.g.\ }               
\def\Eh{\E_\idty}             
\def\chain#1{{#1}^\Irs\ }
\def\om{\omega}
\def\Om{\Omega}
\def\bom{{\overline\omega}}
\def\up#1{^{(#1)}}
\def\qg{$(\C,\copr)$} 
\def\copr{\Delta}
\def\counit{\epsilon}
\def\antipd{\kappa}
\def\multip{{\bf m}}
\def\otimesmin{\otimes_{\mkern-4mu\rm min}}
\def\Csub{\C_0}
\def\extdact{{\widehat\alpha}}
\def\Rh{{\widehat R}}
\def\Sh{{\widehat S}}
\def\Rest{{\bf R}}
\def\Aql{\A_{\rm ql}}

\def\Csubb{\C_1}
\def\M#1#2{{\cal M}_{#1}(#2)}
\def\SnU#1{{\rm S}_\nu{\rm U}(#1)}
\def\SU#1{{\rm SU}(#1)}
\def\Sn2{${\rm S}_\nu{\rm U}(2)$}
\def\q{{\nu}}
\def\j{{J}}

\def\Cuntz{{\cal O}}

\catcode`@=11
\def\ifundefined#1{\expandafter\ifx\csname
                        \expandafter\eat\string#1\endcsname\relax}
\def\atdef#1{\expandafter\def\csname #1\endcsname}
\def\atedef#1{\expandafter\edef\csname #1\endcsname}
\def\atname#1{\csname #1\endcsname}
\def\ifempty#1{\ifx\@mp#1\@mp}
\def\ifatundef#1#2#3{\expandafter\ifx\csname#1\endcsname\relax
                                  #2\else#3\fi}
\def\eat#1{}
\newcount\refno \refno=1
\def\labref #1 #2 #3\par{\atdef{R@#2}{#1}}
\def\lstref #1 #2 #3\par{\atedef{R@#2}{\number\refno}
                              \advance\refno by1}
\def\txtref #1 #2 #3\par{\atdef{R@#2}{\number\refno
      \global\atedef{R@#2}{\number\refno}\global\advance\refno by1}}
\def\doref  #1 #2 #3\par{{\refno=0
     \vbox {\everyref \item {\reflistitem{\atname{R@#2}}}
            {\d@more#3\more\@ut\par}\par}}\vskip\refskip }
\def\d@more #1\more#2\par
   {{#1\more}\ifx#2\@ut\else\d@more#2\par\fi}
\let\more\relax
\let\everyref\relax  
\newdimen\refskip  \refskip=\parskip
\let\REF\labref
\def\@cite #1,#2\@ver
   {\eachcite{#1}\ifx#2\@ut\else,\hskip0pt\@cite#2\@ver\fi}
\def\citeform#1{\lbrack{\bf#1}\rbrack}
\def\cite#1{\citeform{\@cite#1,\@ut\@ver}}
\def\eachcite#1{\ifatundef{R@#1}{{\tt#1??}}{\atname{R@#1}}}
\def\defonereftag#1=#2,{\atdef{R@#2}{#1}}
\def\defreftags#1, {\ifx\relax#1\relax \let\next\relax \else
           \expandafter\defonereftag#1,\let\next\defreftags\fi\next }
\newdimen\refskip  \refskip=\parskip
\def\@utfirst #1,#2\@ver
   {\author#1,\ifx#2\@ut\afteraut\else\@utsecond#2\@ver\fi}
\def\@utsecond #1,#2\@ver
   {\ifx#2\@ut\andone\author#1,\afterauts\else
      ,\author#1,\@utmore#2\@ver\fi}
\def\@utmore #1,#2\@ver
   {\ifx#2\@ut\and\author#1,\afterauts\else
      ,\author#1,\@utmore#2\@ver\fi}
\def\authors#1{\@utfirst#1,\@ut\@ver}
\def\citeform#1{{\bf\lbrack#1\rbrack}}
\let\everyref\relax            
\let\more\relax                
\let\reflistitem\citeform
\catcode`@=12
\def\Bref#1 "#2"#3\more{\authors{#1}:\ {\it #2}, #3\more}
\def\Gref#1 "#2"#3\more{\authors{#1}\ifempty{#2}\else:``#2''\fi,
                             #3\more}
\def\Jref#1 "#2"#3\more{\authors{#1}:``#2'', \Jn#3\more}
\def\inPr#1 "#2"#3\more{in: \authors{\eds#1}:``{\it #2}'', #3\more}
\def\Jn #1 @#2(#3)#4\more{{\it#1}\ {\bf#2}(#3)#4\more}
\def\author#1. #2,{#1.~#2}
\def\sameauthor#1{\leavevmode$\underline{\hbox to 25pt{}}$}
\def\and{, and}   \def\andone{ and}
\def\noinitial#1{\ignorespaces}
\let\afteraut\relax
\let\afterauts\relax
\def\etal{\def\afteraut{, et.al.}\let\afterauts\afteraut
           \let\and,}
\def\eds{\def\afteraut{(ed.)}\def\afterauts{(eds.)}}
\catcode`@=11
\newcount\eqNo \eqNo=0
\def\lasteq{\secNo.\number\eqNo}
\def\deq#1(#2){{\ifempty{#1}\global\advance\eqNo by1
       \edef\n@@{\lasteq}\else\edef\n@@{#1}\fi
       \ifempty{#2}\else\global\atedef{E@#2}{\n@@}\fi\n@@}}
\def\eq#1(#2){\edef\n@@{#1}\ifempty{#2}\else
       \ifatundef{E@#2}{\global\atedef{E@#2}{#1}}%
                       {\edef\n@@{\atname{E@#2}}}\fi
       {\rm(\n@@)}}
\def\deqno#1(#2){\eqno(\deq#1(#2))}
\def\deqal#1(#2){(\deq#1(#2))}
\def\eqback#1{{(\advance\eqNo by -#1 \lasteq)}}

\def\eqgroup(#1){{\global\advance\eqNo by1
       \edef\n@@{\lasteq}\global\atedef{E@#1}{\n@@}}}
\outer\def\iproclaim#1/#2/#3. {\vskip0pt plus50pt \par\noindent
     {\bf\dpcl#1/#2/ #3.\ }\begingroup \interlinepenalty=250\lessblank\sl}
\newcount\pcNo  \pcNo=0
\def\lastpc{\number\pcNo} 
\def\dpcl#1/#2/{\ifempty{#1}\global\advance\pcNo by1
       \edef\n@@{\lastpc}\else\edef\n@@{#1}\fi
       \ifempty{#2}\else\global\atedef{P@#2}{\n@@}\fi\n@@}
\def\pcl#1/#2/{\edef\n@@{#1}%
       \ifempty{#2}\else
       \ifatundef{P@#2}{\global\atedef{P@#2}{#1}}%
                       {\edef\n@@{\atname{P@#2}}}\fi
       \n@@}
\def\Def#1/#2/{Definition~\pcl#1/#2/}
\def\Thm#1/#2/{Theorem~\pcl#1/#2/}
\def\Lem#1/#2/{Lemma~\pcl#1/#2/}
\def\Prp#1/#2/{Proposition~\pcl#1/#2/}
\def\Cor#1/#2/{Corollary~\pcl#1/#2/}
\def\Exa#1/#2/{Example~\pcl#1/#2/}
\font\sectfont=cmbx10 scaled \magstep2
\def\secNo{00}
\def\Beginsection#1#2{\vskip\z@ plus#1\penalty-250
  \vskip\z@ plus-#1\bigskip\vskip\parskip
  \leftline{\bf#2}\nobreak\smallskip\noindent}
\def\bgsection#1. #2\par{\Beginsection{.3\vsize}{\sectfont#1.\ #2 }%
            \def\secNo{#1}\eqNo=0}
\def\bgssection#1. #2\par{\Beginsection{.3\vsize}{#1.\ #2 }%
            \def\secNo{#1}\eqNo=0}
\def\Acknow#1\par{\ifx\REF\doref
     \Beginsection{.3\vsize}{\sectfont Acknowledgements}%
#1\par
     \Beginsection{.3\vsize}{\sectfont References}\fi}
\catcode`@=12
\def\class#1 #2*{{#1},}
\defreftags
AF=AFri, AKLT=AKLT, ASW=ASW, BS=Baaj, Bab=BAB, BMNR=BMNR, BY=BY,
BF=Bernard, Ber=Bernard2, Bie=Biedenharn, BR=BraRo, CE=CEffros,
Cu1=Cuntz, Cu2=Cuntzact, DC=Dasgupta, DFJMN=DFJMN, DHR=DHR,
DR1=Roberts, DR2=Roberts2, Dri=Drinfeld, DW=LMD, FNW1=FCS, FNW2=FCP,
FSV=FSV, GW=GW, GS=Grosse, Ha1=Haldane, Ha2=Haldanereview, Jim=Jimbo,
JSW=WICK, KSZ=Zitt, KNW=Watatani, KS=Kulish, LB=Ma,
MS=Mack, Maj=Majid, Mat=Mathematica, MMP=Zagreb, MN=MezNep, RW=MF,
Rue=Ruegg, Sha=Shastri, SV=SV, Tak=Takesaki, Vec=Vecser, Wo1=WoRims,
Wo2=Woronowicz, Wo3=WoronDC, Wo4=WoronCG, ,

\font\BF=cmbx10 scaled \magstep 3
\line{{\tt cond-mat/9504002}\hfill Preprint KUL-TF-94/8}
\line{\hfill to appear in \it Commun.Math.Phys.}
\voffset=2\baselineskip
\hrule height 0pt
\vskip 40pt plus40pt
\centerline{{\BF Quantum Spin Chains with}}
\vskip10pt
\centerline{{\BF  Quantum Group Symmetry}}
\vskip 30pt plus30pt
\centerline{
  M.~Fannes$^{1,2}$, B.~Nachtergaele$^{3,4}$, and R.F.~Werner$^5$}
\vskip 20pt plus40pt
\noindent {\bf Abstract}\hfill\break
{We consider actions of quantum groups on lattice spin systems.
We show that if an action of a quantum group respects the local
structure of a lattice system, it has to be an ordinary group.
Even allowing weakly delocalized (quasi-local) tails of the action,
we find that there are no actions of a properly quantum group
commuting with lattice translations.
The non-locality arises from the ordering of factors in the
quantum group C*-algebra, and can be made one-sided, thus allowing
semi-local actions on a half chain.
Under such actions, localized quantum group invariant elements remain
localized. Hence the notion of interactions invariant under the
quantum group and also under translations, recently studied by many
authors, makes sense even though there is no global action of the
quantum group.
We consider a class of such quantum group invariant interactions
with the property that there is a unique translation invariant
ground state.
Under weak locality assumptions, its GNS representation carries no
unitary representation of the quantum group.
}

\noindent {\bf Mathematics Subject Classification (1991):}
\hfill\break
\class 81R50   Quantum groups and related algebraic methods*
\class 82B10   Quantum equilibrium statistical mechanics
                   (general)*
\class 46L60   Applications of selfadjoint operator algebras to
               physics*

\def\Email{\nl\vrule width0pt\qquad Email: \tt }
\vfootnote1
  {Inst. Theor. Fysica, Universiteit Leuven, B-3001 Heverlee,
   Belgium.
   \Email mark.fannes@fys.kuleuven.ac.be}
\vfootnote2
  {Onderzoeksleider, N.F.W.O. Belgium}
\vfootnote3
  {Dept. of Physics, Princeton University, NJ-08544-708, USA;
  \Email bxn@math.princeton.edu  }
\vfootnote4
  {Supported in part by NSF Grant \# PHY90-19433 A02}
\vfootnote5
  {Fachbereich Physik, Universit\"at Osnabr\"uck,
  49069 Osnabr\"uck, Germany.
  \Email reinwer@dosuni1.rz.uni-osnabrueck.de}
\vfill\eject

\voffset=0.0truecm
\bgsection 1. Introduction

Symmetry has always played an important role in theoretical physics
in helping to reduce a problem with many variables to a more tractable
size. In statistical mechanics we have infinitely many degrees of freedom
to deal with, so often the symmetry, while helpful, is not sufficient
to solve the problem, unless we have ``infinitely many symmetries''.
One example is the theory of mean-field lattice systems, where the
inherent permutation symmetry is sufficient to reduce the computation
of limit free energy density, of the possible limit states
\cite{FSV,MF}, and of the limit dynamics \cite{LMD} to corresponding
problems in the algebra for a single spin.

Another example, which has been studied intensively by many authors
recently \cite{BAB,BMNR,Dasgupta,Kulish,MezNep,Grosse}, is the class
of models which can be solved exactly (though not always rigorously)
by means of the Bethe Ansatz. The basis of this method is to
diagonalize the Hamiltonian along with an infinite set of constants
of motion. In some cases the occurrence of this infinite set of
constants of motion is related to the appearance of a new kind of
symmetry, called quantum group symmetry. This nourishes the hope
that by relaxing the demands usually made on the structure of a
symmetry group, and allowing the wider class of quantum groups, one
can benefit from symmetry considerations in new situations, where a
symmetry in the traditional sense is simply not present.
A particularly interesting development in this direction is the
integrable Haldane-Shastri model \cite{Haldane,Shastri}. This quantum
spin chain with long-range interactions can be interpreted as an ideal
semion gas, i.e. the spinon excitations obey fractional statistics.
For more details and generalizations we refer the reader to
\cite{Haldanereview} and the references therein. Another example
where a quantum group symmetry plays an explicit role is in the
study of non-translation invariant ground states of the ferromagnetic
XXZ chain \cite{GW,ASW}.

In this paper we exclusively consider one-dimensional
quantum spin systems, or ``spin chains'',
as a testing ground for applications of quantum group
symmetries. We emphasize that here we use the word ``spin chain'' in
its meaning familiar from statistical mechanics, i.e.\ spins at
different sites commute, and we do not consider modified (braided)
tensor products \cite{Majid}.
A number of models of this type have been
considered in the recent literature \cite{Zitt,BY}.
The interactions in these models are both translation invariant, like
the usual lattice interactions, and quantum group invariant in a
sense we will make more precise below. It is thus natural to ask for
the quantum group symmetry of the relevant states --- temperature
and ground states--- of these models. For example, could there be
``spontaneous quantum symmetry breaking''? In the case of ordinary
groups it is clear how to define such notions: the symmetry is
implemented locally by unitaries, in a way which is compatible with
the thermodynamic limit. The symmetry group thus acts by
automorphisms on the infinite system described by the quasi-local
algebra, and it is with respect to this action that we can talk
about ``invariant states'' of the infinite system.

As we will show in this paper, however, this approach does not work
for quantum group symmetries. The first limitation is that the
formation of tensor products with non-abelian coefficients requires
an ordering of the factors in the product, which is correlated with the
algebraic ordering of factors in the algebra. This limits all
considerations to one dimensional systems, or, in the field
theoretical context, to systems with one space dimension and one time
dimension. If we make the technical simplifications of choosing a
discrete space variable, and a finite dimensional one-site algebra,
we arrive at the setting of quantum spin chains, used in this paper.
Given a unitary action on the one-site algebra, we can define a
product action of a quantum group, for each finite segment of the
chain. The fundamental difficulty, however, is that these actions are
{\it not} compatible with the identifications used to form the
inductive limit to the infinite system. More precisely, the
compatibility holds for enlargement of the system towards the right,
but not towards the left (or, conversely, depending on conventions).
This means that we can define quantum group actions on a
semi-infinite chain, but not on the full chain. We prove that this
restriction is inherent in the quantum group concept, by showing
that there is no action on the quasi-local algebra of the chain
which commutes with translations.

In \cite{DFJMN} Davies et al.\ study the quantum group
symmetries of the anti-ferroelectric XXZ chain. The infinite
dimensional symmetry algebra introduced there contains the finite
dimensional quantum group $\SnU2$. Therefore our results, in
particular \Thm11/qloc/,  imply that the construction of
\cite{DFJMN} cannot lead to a proper action of the quantum
symmetries on the observable algebra which commutes with the
translations of the chain (or any infinite subgroup of the
translations). It is an interesting open question in what sense such
an action could be defined.

In contrast to the actions defined for each segment, the sets of
invariant elements for these actions are compatible with the
inductive limit. This allows us to define quantum group invariant
interactions (see, e.g.\ \cite{Zagreb}), even though this invariance
cannot be understood as invariance with respect to a global action.

The fact that the invariant elements have much better localization
properties than general elements is reminiscent of the theory of
superselection sectors in relativistic quantum field theory.
Interpreting the quantum group as a gauge group, one would consider
only the invariant elements as ``observables''. The rest of the
algebra would then be an algebra of unobservable fields, whose
function in the theory is to describe operations changing the
superselection sector (``creating a charge''). Already in the case of
Fermi fields one has to relax the requirement of locality for the
field algebra (allowing anti-commutation rules). The lack of locality
for the action of the quantum group on the ``fields'' could be seen as
a reflection of this general feature of field algebras. Indeed, some
constructions of field algebras in this context lead to very non-local
objects \cite{Bernard,Bernard2}.

The theory of superselection sectors in field theory provides some of
the most interesting application of quantum groups. The basic problem
is to relate the set of sectors, together with their composition
("fusion") rules, to the set of irreducible representations of a
group or quantum group with the rules for decomposing tensor
products. In two and more space dimensions this program has been
carried out with complete success by Doplicher and Roberts
\cite{Roberts,Roberts2}, building on earlier work together with Haag
\cite{DHR}. They managed to show, in two or more space dimensions,
and using only axiomatic assumptions on the observable algebra, that
the superselection structure indeed comes from the representation
theory of a compact gauge group. They also reconstructed an algebra
of fields with an action of the gauge group, whose fixed points are
precisely the observables. The fact that they get a (non-quantum)
group depends crucially on having more than one space dimension, and
hence the possibility of exchanging two spacelike regions in a
continuous process during which they always remain spacelike. In one
space dimension the superselection structure can be much more
complex \cite{Mack,SV}. At the same time, there is a rich supply of
explicit models with conformal symmetry, for which the
superselection structure can be computed (see \eg \cite{Vecser}).

Regarding the connection with the present paper, we wish to point
out, however, that quantum groups of the kind we use are not so
interesting for the project of reconstructing superselection
structures. For example, the irreducible representations of the
quantum deformation $\SnU2$ of $\SU2$ and the decomposition weights
for tensor products are precisely the same as for $\SU2$ for real
values of the deformation parameter $\nu$. New features, such as
structures with only finitely many sectors, are seen only for
complex values of $\nu$, particularly roots of unity. In that case,
however, one loses the involution in the algebra of ``functions'' on
the group, and with it the notion of an action on the observable
algebra, which is the object of our investigation.

The decomposition of the algebra of a spin chain with respect to the
representations of a group is also reflected in the decomposition
theory of invariant states. We discuss one possibility of defining
quantum group invariant states even when there is no action: a
``hereditarily invariant state'' has the property that all its
restrictions to finite segments are invariant to the quantum group
action given for the segment. Unfortunately, this seems to give no
interesting result: on the basis of computations on short chains we
conjecture that for $\SnU2$ only one state (an infinite product
state) has this property.

In the case of classical groups, there is a general construction
\cite{FCS,FCP} yielding non-trivial states on a chain which are
translationally invariant, and also invariant under the action of the
group. These states are automatically ground states of a suitable
finite range interaction. The whole construction is naturally
covariant, also with respect to quantum groups. It then yields a class
of translationally invariant finite range interactions, which are also
invariant with respect to an irreducible representation of a quantum
group, and have a unique ground state (see \cite{Zitt} for the
minimal non-trivial example in this class). Of course, in the case of
a classical group the unique ground state is then also invariant under
the group. Not so for quantum groups: on the full chain we cannot even
say what an invariant state should be, because there is no action of
the quantum group. On the half chain, where we can define an action of
the quantum group, the uniqueness of the ground state fails, and we
get a finite dimensional set of ground states, parametrized by a
boundary condition. Among these ground states we now have one state
which is translationally invariant, and another state, which is quantum
group invariant. Of course, the two are different.

Since our main objective is to point out the difficulties in
combining local structure with quantum group symmetry, we have not
aimed at maximum generality. The only concrete quantum group we
consider is Woronowicz's one-parameter deformation $\SnU2$ of
$\SU2$. Since this example has served as the paradigm of a quantum
group in many papers, we are confident that the difficulties pointed
out by us are indeed typical. There are two properties which we
prove for $\SnU2$ (\Prp4/SUq/ and \Prp5/P.non-ext/ ) which can be
stated for general quantum groups. We would like to pose their
generalization to other quantum groups as a challenge to experts in
the field. Even in case of $\SnU2$ we had to leave unsettled one
statement (Conjecture \pcl8/conject/), which implies, among other
things, that only a specific product state is both quantum group and
translationally invariant for the canonical action on the half
chain.  Much of the literature is phrased in terms of quantum groups
in the sense of Drinfel'd rather than Woronowicz. We chose the
latter definition because the notion of ``action'' seemed more
natural in this context. The connection between the two approaches
is briefly indicated in the Appendix.

The paper is organized as follows. In Section 2 we review briefly the
notion of quantum group in the sense of Woronowicz
\cite{Woronowicz,WoronCG}, and of the action of a quantum group on a
C*-algebra, of fixed points under such an action, and of invariant states
with respect to such an action.
In Section 3 we consider the operation of tensor product for unitary
representations and for actions, and describe the basic locality
problem for such tensor products.  We introduce a more restrictive definition
of ``actions'', which seems more natural for discussing tensor
products. Unfortunately, where the standard definition leads to
locality problems for extending the chain to the left, the more
restrictive definition creates problems right and left.
In Section 3, we also define the action on a half
chain associated with a unitary representation on the one-site
Hilbert space, and its construction in terms of an action on the
Cuntz algebra $\Cuntz_d$.
In Section 4, we define invariant elements, and show the
compatibility of this notion with the local structure, and discuss
the hereditarily invariant states.
Section 5 contains the NO-GO Theorem for actions on the quasi-local
algebra, and Section 6 discusses similar problems for the
implementation of actions by unitaries in the GNS representation of
an invariant state.
In Section 7, we discuss the quantum group covariance of the
construction of finitely correlated states.
The decay rate in such a
state is given by a quantum Wigner $6j$-symbol. These have been
computed in detail \cite{Ruegg,Biedenharn,Ma},
so, in principle, we can save ourselves the work of diagonalizing a
transfer operator.
In the explicit example of the q-AKLT model. however,
diagonalizing the transfer matrix directly by hand is so
straightforward that checking the conventions used in any particular
computation of Wigner $6j$-symbols would not be worth the effort.

In order to do some of the more tedious quantum group computations
reliably, we developed a package for {\tt Mathematica}
\cite{Mathematica}, which is available by anonymous {\tt ftp} from
{\tt nostromo.physik.Uni-Osnabrueck.de}.\par

\bgsection 2. Quantum groups

As there is not yet a standard notion of quantum group (also called
pseudogroup) and of the related invariance and covariance
properties, we will briefly review how quantization works for the
case of compact groups $\G$. Our discussion will be based completely
on the notion of quantum groups introduced by Woronowicz
\cite{Woronowicz,WoronCG,WoRims,WoronDC}. An alternative would be the
Drinfel'd approach \cite{Drinfeld,Jimbo}, which provides a
``quantization''  of Lie algebras rather than groups. Some of the
questions considered in this paper could also be posed using this
approach, but we found the Woronowicz approach more suited for this
purpose. On the other hand, the Drinfel'd approach is much more
effective for doing explicit computations. Therefore, for the
reader's convenience, we have included a brief Appendix on the
connection of these approaches. A new approach to quantum groups has
recently been initiated by Baaj and Skandalis \cite{Baaj}. In this
context actions on C*-algebras have been considered by Cuntz
\cite{Cuntzact}.

The topology of a compact group $\G$ is encoded in the algebra $\C(\G)$ of
continuous, complex-valued functions on the group. This is a $\ast$-algebra
under the natural notions of addition, multiplication and complex
conjugation. Equipped with the supremum norm, $\C(\G)$ becomes a
commutative C*-algebra with identity. That this algebra carries the
complete information about $\G$ as a topological space, is the content of
the ``Gel'fand Isomorphism Theorem'' which reconstructs, starting from any
commutative C*-algebra, the compact space on which this algebra is the
algebra of continuous functions. The next step is to encode the
multiplication operation of $\G$. Three maps are naturally connected to the
composition law in $\G$, the existence of a neutral element $e\in\G$ and of
the inverse $g^{-1}$ of any $g\in\G$, respectively. These three maps
become, in turn,
\begingroup\lessblank
\item{i)}
  the {\it coproduct\/} $\copr$ which maps $\C(\G)$ into
  $\C(\G\times\G)\cong\C(\G)\ot\C(\G)$, the complex continuous functions in
  two variables:
  $$(\copr \,f)(g_1,g_2)= f(g_1\,g_2), \qquad f\in\C(\G),\quad
  g_1,g_2\in\G.$$
\item{ii)}
  the {\it antipode\/} $\antipd$ which maps $\C(\G)$ into itself, given by:
  $$\antipd(f)(g)= f(g^{-1}), \qquad f\in\C(\G),\quad g\in\G.$$
\item{iii)}
  the {\it counit\/} $\counit$ which is the character
  $$\counit(f)=f(e) \qquad f\in\C(\G).$$

\noindent
The group--axioms are reflected in the properties of the maps $\copr$,
$\antipd$ and $\counit$:
\item{i)}
  associativity of the composition law in $\G$:
  $$(\copr\ot\id)\circ\copr = (\id\ot\copr)\circ\copr
\deqno(co-ass)$$
\item{ii)}
  $e$ is the neutral element in $\G$:
  $$(\counit\ot\id)\circ\copr = (\id\ot\counit)\circ\copr =
  \id \deqno(co-neut)$$
\item{iii)}
  $g^{-1}$ is the inverse of $g$ in $\G$:
  $$\multip\left((\id\ot\antipd)\circ\copr\right) =
  \multip\left((\antipd\ot\id)\circ\copr\right) =
  \counit\idty, \deqno(co-inv)$$
  where $\multip$ is the {\it multiplication map\/} from
  $\C(\G)\ot\C(\G)\to\C(\G)$ taking $f\ot g$ into $fg$.
\endgroup

\noindent
We could now consider an abelian algebra that comes with such maps $\copr$,
$\antipd$ and $\counit$ and reconstruct the compact group $\G$. The key
point is, however, that we have nowhere used the commutativity of $\C(\G)$,
so we can drop this assumption, and arrive at the more general notion of
quantum groups. In dropping the commutativity assumption problems arise
with the boudedness of $\antipd$, $\counit$ and $\multip$. The
multiplication map $\multip$ on $\B(\H)\ot\B(\H)$, for instance, has norm
$\dim\H$ (consider the unitary flip operator
$F\phi\otimes\psi=\psi\otimes\phi$ on $\B(\H)\otimes\B(\H)$, for which
$\norm{\multip(F)}=\dim\H$). The following definition, due to Woronowicz
\cite{WoronCG}, takes care of this difficulty:

\iproclaim/Dqg/ Definition.
A {\bf compact quantum group} \qg\ consists of:
\item{i)}
  a separable C*-algebra $\C$ with identity $\idty$ and
\item{ii)}
  a unital *-homomorphism $\copr: \C\to\C\otimesmin\C$.

\noindent
such that
\item{i)}
  $(\copr\ot\id)\circ\copr = (\id\ot\copr)\circ\copr$, and
\item{ii)}
  both $\copr(\C)(\idty\ot\C)$ and $\copr(\C)(\C\ot\idty)$ are dense in
  $\C\otimesmin\C$.
\eproclaim

It is shown in \cite{WoronCG} that there exists a dense $\ast$-subalgebra
$\C_0$ of $\C$ such that $\copr(\C_0)\subset\C_0\odot\C_0$, where $\odot$
denotes the algebraic tensor product of $\C_0$ with itself \ie the finite
linear combinations of elements of the form $a\otimes b$, $a,b\in\C_0$.
Furthermore $\C_0$ is a Hopf $\ast$-algebra. This means that there are,
uniquely determined maps, $\antipd$ and $\counit$, such that:
{\lessblank
\item{i)}
  $\antipd$ is a
  lin\-ear, anti\-multi\-plic\-ative map from $\C_0$ into it\-self
  that sat\-is\-fies \hfill\break
  $\antipd\bigl(\bigl(\antipd(a^*)\bigr)^*\bigr)=a$,
  $a\in\C_0$ and
\item{ii)}
  $\counit$ is a $\ast$-preserving character on $\C_0$.

\noindent}
The maps $\copr$, $\antipd$ and $\counit$ satisfy the equations
\eq(co-ass), \eq(co-neut) and \eq(co-inv).
The dense $\ast$-subalgebra $\C_0$ consists of all matrix elements of the
finite-dimensional unitary representations of \qg\ (the notion of unitary
representation will be introduced shortly).

The standard example of such a structure is the one-parameter deformation
$\SnU2$ of $\SU2$. Such a deformation is rather drastic in so far that the
commutative algebra of complex functions on $\SU2$ is replaced by an algebra
$\C$ with trivial center (see \Prp4/SUq/). Still, the whole representation
theory of
$\SnU2$ turns out to depend smoothly on $\nu$.

\noindent
{\it Example:} \nl
Let $-1\leq\nu\leq1$ and $\C$ be the C*-algebra with unit, generated by
$\alpha$ and $\gamma$, which satisfy the relations:
$$\eqalign{\alpha\alpha^*+\nu^2\gamma^*\gamma &= \idty \cr
           \alpha^*\alpha+\gamma\gamma^* &= \idty \cr
           \alpha\gamma^*-\nu\gamma^*\alpha &= 0 \cr
}\deqno(Snu2)$$
The relations
$$\gamma\gamma^*=\gamma^*\gamma \midbox{and}
\alpha\gamma= \nu\gamma\alpha$$
follow automatically \cite{WICK}. The coproduct, antipode and counit
are determined by:
$$\eqalign{ \copr\alpha = \alpha\ot\alpha- \nu\gamma^*\ot\gamma
&\midbox{} \copr\gamma = \gamma\ot\alpha + \alpha^*\ot\gamma \cr
\antipd(\alpha) = \alpha^* &\midbox{}
\antipd(\gamma) = -\nu\gamma \cr
\counit(\alpha) = \idty &\midbox{}
\counit(\gamma) = 0. \cr
}\deqno(Snu2op)$$
The relations between $\alpha$ and $\gamma$ are such that
$$
  u=\left(\matrix{\alpha&-\nu\gamma^*\cr
                  \gamma&\alpha^*}\right)
$$
is a unitary in ${\cal M}_2\ot\C$. This $u$ is called the {\it fundamental
representation\/} of $\SnU2$.

In order to define representations of quantum groups and to construct
products of representations two new products are introduced. Let $\A$, $\B$
and $\C$ be C*-algebras with units. We put for $A\in\A$, $B\in\B$ and
$C_1,C_2\in\C$:
$$\eqalign{
(A\ot C_1)\tee(B\ot C_2)
&= A\ot B\ot C_1C_2 \cr
&= (A\ot C_1\ot\idty_\B)\,(\idty_\A\ot B\ot C_2)
\quad,}\deqno(Dtee)$$
and, for $A_1,A_2\in\A$ and $C_1,C_2\in\C$:
$$(A_1\ot C_1)\eet(A_2\ot C_2) = A_1A_2\ot C_1\ot C_2
\quad.\deqno(Deet)$$
It should be stressed that both $\tee$ and $\eet$ involve an ordinary
product in a non-commutative algebra. Therefore the order of the factors is
quite important and also the $\ast$-operation will behave badly with
respect to these products. Although the the multiplication map
$\multip:\C\ot\C\to\C$ is not bounded on an infinite dimensional
non-abelian algebra $\C$, the norm estimate $\norm{X\tee
Y}\leq\norm{X}\norm{Y}$ holds.

We can now define the analogues of many concepts of classical
group theory. In each case it is easy to verify that for abelian $\C$,
that is for an ordinary group, the new concept coincides with the
ordinary one. When there is a possibility of confusion, we will
denote the unit element of an algebra $\A$ by $\idty_\A$, and the
identity map on $\A$ by $\id_\A$.

\titem
A {\it unitary representation} $v$ of a quantum group \qg\
on a Hilbert space $\H$ is a unitary element
$v\in\B(\H)\otimes\C$ such that $v\eet v= (\id\ot\copr)(v)$.
Suppose that $\H$ is $k$-dimensional and let
$\set{f_{ij}\stt i,j=1,2,\ldots k}$ be
matrix units in $\M k\Cx$. $v$ can then be written as:
$$v = \sum_{ij} f_{ij}\ot v_{ij}, \quad v_{ij}\in\C
\quad.\deqno(matunit)$$
We can thus consider the $v_{ij}$ as the matrix elements of a $\C$-valued
matrix, and identify $\M k\Cx\otimes\C$ with $\M k\C$, the $\C$-valued
$k\times k$-matrices. In terms of the
$v_{ij}$ the representation condition is:
$$\copr(v_{ij})= \sum_{\ell=1}^k v_{i\ell}\ot v_{\ell j}
\quad.\deqno(repvij)$$

\titem
A linear operator $W:\H_1\to\H_2$ {\it intertwines} between the unitary
representations $v_1$ and $v_2$ of \qg\ on $\H_1$ and $\H_2$ if
$(W\ot\idty_\C)\, v_1= v_2\, (W\ot\idty_\C)$.

\titem
A unitary representation $v$ of \qg\ is {\it irreducible} if the only
intertwiners between $v$ and $v$ are the multiples of the identity.

\titem
A state $h$ on $\C$ is called a {\it Haar measure} if:
$$(h\ot\id)\circ\copr= (\id\ot h)\circ\copr= h
\quad.\deqno(Hair)$$
In this formula the right hand side is to be read as the map taking
$a\in\C$ to $h(a)\idty_\C$.

\titem
A unitary representation $v$ of \qg\ on $\H$ {\it implements} an
action $\alpha_v$ of \qg\ on $\B(\H)$ by restricting $\ad(v)$ to
$\B(\H)\ot\idty_\C$:
$$ \alpha_v(A)= \ad(v)(A\ot\idty_\C)
              = v(A\ot\idty_\C)v^*
\quad,\qquad A\in\B(\H)
\quad.\deqno(adv)$$
More generally, an {\it action} of a quantum group \qg\ on a C*-algebra
$\A$ is a $\ast$-homomorphism $\alpha$ of $\A$ into $\A\otimesmin\C$
mapping the identity of $\A$ into that of $\A\ot\C$ and such that:
$$(\alpha\ot\id_\C) \circ \alpha
= (\id_\A\ot\copr) \circ \alpha \quad.\deqno(action)$$

\titem
We will say that a state $\om$ on $\A$ is {\it invariant} under an
action $\alpha$ of \qg\ on $\A$ if for all $A\in\A$,
$(\om\ot\id_\C)(\alpha(A)) = \om(A)\idty_\C$.

\noindent
Let us show the existence of invariant states for a quantum group with a
Haar measure $h$ acting on a C*-algebra $\A$ by $\alpha$.
For any state $\om$ on $\A$, define the average $\bom$ over the group by:
$\bom(A) = (\om\ot h)(\alpha(A))$, $A\in\A$. A simple computation
shows that $\bom$ is $\alpha$-invariant:
$$\eqalign{(\bom\ot\id_\C)(\alpha(A))
&= (\om\ot h\ot\id_\C) \bigl((\alpha\ot\id_\C)(\alpha(A))\bigr) \cr
&= (\om\ot h\ot\id_\C) \bigl((\id_\A\ot\copr)(\alpha(A))\bigr) \cr
&= (\om\ot (h\ot\id_\C)\circ\copr)(\alpha(A)) \cr
&= (\om\ot h)(\alpha(A)) \cr
&= \bom(A).
}$$

Unitary representations and actions on C*-algebras are special cases
of ``{\it linear representations} on a vector space''. Yet the
definitions look slightly different: we took a unitary representation
as an element $v\in\B(\H)\otimes\C$, and an action as a map
$\alpha:\A\to\A\otimes\C$. The classical intuition for all
representations is that under an action (or representation) $R$ the
vector $x\in X$ becomes a function on the group with values in $X$,
i.e.\ an element of $X\otimes\C$. Thus a representation is a map
$R:X\to X\ot\C$, and the compatibility with the product becomes
encoded in the relation
$$ (R\ot\id_\C)\circ R
      =   (\id_X\otimes\copr)\circ R
\quad.\deqno(linrep)$$
Of course, when $X$ is finite dimensional, we can set
$Re_i=\sum_je_j\otimes R_{ji}$, where $R_{ji}\in\C$ satisfy
\eq(repvij). This is the $\C$-valued matrix we used for the
definition of unitary representations. The difference between unitary
representations on Hilbert spaces and actions on C*-algebras is thus
mainly in the structure of the underlying space and the sense in
which it is preserved by the representation: unitarity is most
conveniently formulated in terms of $v\in\B(\H)\otimes\C$, whereas
the homomorphism property is more easily expressed in terms of
$\alpha:\A\to\A\otimes\C$.

For the definition of tensor products it is important to apply a
representation not only to $X$ (i.e.\ to ``group independent
vectors'', but also to vectors $x\in X\otimes\C$ which already depend
on a group element. Thus we also need to consider maps
$$ \Rh:X\otimes\C\to X\otimes\C
\quad.$$
In the classical case, when $\C=\C(G)$, we can define $\Rh$ in
terms of $R$. In order to do this, we identify $X\otimes\C$ with the
algebra of $X$-valued continuous functions on $G$, and set, for
continuous $x:G\to X$,
$$ (\Rh x)(g)=R_g\bigl(x(g)\bigr)
\quad.\deqno(repex)$$
More abstractly, this can be written as
$$ \Rh (x\otimes C)=R(x)\, \id_X\otimes C
\quad,\deqno(repexx)$$
where the product on the left is shorthand for
$(x\otimes C')(\id_X\otimes C)=x\otimes(C'C)$. Equation \eq(repex)
makes sense in the quantum group case as well,
and one readily verifies that the representation relation for
$\Rh$ becomes
$$(\Rh\ot\id_2) \circ (\id_1\ot\Rh) \circ (\id_\A\ot\copr)
    = (\id_\A\ot\copr) \circ \Rh
\quad.\deqno(actionRh)$$
This is an equation between maps $X\ot\C\to X\ot\C\ot\C$, and the
subscripts 1 and 2 of the identity maps refer to the first and second
tensor factor $\C$. Tensor factors $X$ and $\C$ have to be reshuffled
but the order of the $\C$ factors is kept unchanged.

{From} these considerations it seems that the view of an action as a map
$\Rh$ on $X\otimes\C$ satisfying \eq(actionRh) is simply equivalent to
the general definition in equation \eq(linrep). However, this is true
only as long as we do not consider additional structures on $X$: a
representation on a Hilbert space $X$ is required to be {\it
unitary}, and a representation on a C*-algebra $X$ is required to be
a {\it homomorphism}. We have seen that for a unitary representation
$R$ we can always pass from $R$ to $\Rh$ by \eq(repexx). However,
for actions on a C*-algebra this choice of $\Rh$ destroys the
homomorphism property. Therefore, the following definition is needed
to single out the good cases.

\iproclaim/D.exdact/ Definition.
An {\bf extended action} of a quantum group \qg\ on a C*-algebra $\A$
is an automorphism $\extdact$ of $\A\otimesmin\C$ such that
$$(\extdact\ot\id_2) \circ (\id_1\ot\extdact) \circ (\id_\A\ot\copr)
    = (\id_\A\ot\copr) \circ \extdact
\quad.\deqno(actionp)$$
An action $\alpha:\A\to\A\otimesmin\C$ is called {\bf extendible}, if
it is the restriction of an extended action to $\A\ot\idty_\C$.
\eproclaim

One easily verifies that, for any unitary representation $v$, $\ad(v)$
is an extended action, hence any implemented action in the sense of
the above definitions is automatically extendible. Of course, any
action of a classical group is also extendible. It is not immediately
obvious, then, that there are non-extendible actions at all. However, we
will give an example below, in \Prp5/P.non-ext/, showing that
\Def/D.exdact/ has non-trivial content.

\vfill\eject
\bgsection 3. Tensor products of representations and actions

In this paper we are mainly interested in the action of quantum groups
on composite quantum systems, i.e.\ in actions on a tensor product.
Let $\alpha$ and $\beta$ be actions of a quantum group \qg\ on $\A$
and $\B$ respectively. The product of $\alpha$ and $\beta$ should then
be a homomorphism from $\A\ot\B$ into $\A\ot\B\ot\C$. Let $B\in\B$.
For a non-trivial action, $\beta(B)$ will have components in $\C$ and,
as we only know how to act with $\alpha$ on elements of the form
$A\ot\idty_\C$, we cannot apply $\alpha\ot\id_\B$ to $A\ot\beta(B)$,
$A\in\A$ and $B\in\B$. Therefore the general notion of action as
defined in \eq(action) is ill-adapted to tensor constructs.

It is clear from the discussion at the end of the previous section
what is missing: we need to define actions as operators on
$\A\otimes\C$. With this modified definition of actions it is clear
how to define tensor products of general representations: let
$\Rh:X\otimes\C\to X\otimes\C$ and $\Sh:Y\otimes\C\to
Y\otimes\C$ be ``extended representations'' in the
sense of equation \eq(actionRh). Then we set
$$ \Rh\tee \Sh
    =(\Rh\otimes\id_Y)(\id_X\otimes\Sh)
\quad,\deqno(actensor)$$
with the obvious reshuffling of tensor factors. One then verifies
that $\Rh\tee \Sh$ is indeed again an extended representation.
Moreover, it is obvious that if $\Rh$ and $\Sh$ are both
$\ast$-homomorphisms, or unitary, then so is their $\tee$-product.
Of course, the definition agrees with the usual tensor product in
the abelian case, provided $R$ is extended to $\Rh$ by virtue of
equation \eq(repex). Note, however,  that $\Rh\tee\Sh$ and
$\Sh\tee\Rh$ differ not only in the order of the factors $X$ and
$Y$, which could be undone by a suitable flip isomorphism, but also
by the ordering of the factors in $\C$.

Of course, we can use the extension \eq(repexx) to extend an arbitrary
representation $R$ to $\Rh$, and thus define
$R\tee S=(\Rh\otimes\id_Y)(\id_X\otimes S)$ for such representations.
This coincides, in fact, with the standard definition of tensor
products of unitary representations. It is unsuitable for actions on
C*-algebras, however, since it would practically never lead to a
homomorphism, and hence not to an action in the sense of \eq(action).

For unitary representations, say a representation
$v\in\B(\H)\otimes\C$, and $w\in\B(\K)\otimes\C$,
the $\tee$-product can be written out in
terms of matrix elements as
$$(v\tee w)_{i\mu,j\nu}= v_{ij}\ w_{\mu\nu}
\quad,\deqno(unitensor)$$
where latin and greek indices run over bases of $\H$ and $\K$,
respectively. Obviously, this use of the symbol ``$\tee$'' is also
consistent with the definition given in \eq(Dtee).

In the sequel, we will always consider actions $\alpha_v$
implemented by a unitary representation $v$. Since such actions are
always extendible ($\widehat\alpha_v=\ad(v)$), their $\tee$-product
is well defined according to \eq(actensor). If $v$ and $w$ are as in
\eq(unitensor), we find
$$\ad(v\tee w) = \ad(v)\tee\ad(w)
\quad.\deqno(adtee)$$
The following Theorem summarizes the ``locality'' properties of the
tensor product of two representations.

\iproclaim/2pts/ Theorem.
Let $v$ and $w$ be unitary representations of a quantum group \qg\ on
Hilbert spaces $\H$ and $\K$. Then
\item{(1)}
$$ \alpha_{v\tee w}(\B(\H)\ot\idty_\K) =
   \alpha_v(\B(\H))\ot\idty_\K
    \subset \B(\H)\ot\idty_\K\ot\C
\quad.$$
\item{(2)}
If $B\in\B(\K)$ is $\alpha_w$-invariant, i.e.\
$\alpha_w(B) = B\ot\idty_\C$, then
$$ \alpha_{v\tee w}(\idty_\H\ot B) =
    \idty_\H\ot B\ot\idty_\C
\quad.$$
\item{(3)}
If $\H$ and $\K$ are finite dimensional and if the quantum group is
$\SnU2$, then, conversely,
$\alpha_{v\tee w}(\idty_\H\ot B)\in\idty_\H\ot\B(\K)\ot\C$
implies that $B\in\B(\K)$ is invariant.
\eproclaim

\proof:
In case (1) we have
$$\eqalign{
  \alpha_{v\tee w}(\A\ot\idty_\K)
  &= (v\tee w)\, (\A\ot\idty_\K\ot\idty_\C)\, (v\tee w)^* \cr
  &= (v\ot\idty_\K)\, (\idty_\H\ot w)\, (\A\ot\idty_\K\ot\idty_\C)
  (\idty_\H\ot w)^*\, (v\ot\idty_\K)^* \cr
  &= (v\, \A\ot\idty_\C\, v^*)\ot\idty_\K \cr
  &\subset (\A\ot\idty_\K\ot\C)
\quad.}$$

In case (2):
$$\eqalign{
  \alpha_{v\tee w}(\idty_\H\ot B)
  &= (v\tee w)\, (\idty_\H\ot B\ot\idty_\C)\, (v\tee w)^* \cr
  &= (v\ot\idty_\K)\, \bigl(\idty_\H\ot (w\, B\ot\idty_\C\,
  w^*)\bigr)\, (v\ot\idty_\K)^* \cr
  &= (v\ot\idty_\K)\, (\idty_\H\ot B\ot\idty_\C)\,
  (v\ot\idty_\K)^* \cr
  &= \idty_\H\ot B\ot\idty_\C
\quad.}$$

(3) It is useful to express the action of $\alpha_{v\tee w}$ in matrix
elements with respect to some bases in $\Cx^d$ and $\Cx^k$. Then $v\in\M
d\C$ has matrix elements $v_{ij}\in\C$, $i,j=1,\ldots,d$, and
$w\in\M k\C$ has matrix elements $w_{\lambda\mu}\in\C$, $\lambda,
\mu=1,\ldots,k$. Then
$$\eqalign{
 \Bigl( \alpha_{v\tee w}(A\ot B)\Bigr)_{i\lambda,i'\lambda'}
   &= \Bigl((v\tee w)(A\otimes B\otimes \idty_\C)
        (v\tee w)^*\Bigr)_{i\lambda,i'\lambda'} \cr
   &=\sum_{j\mu j'\mu'}
         v_{ij}w_{\lambda\mu}\
         A_{jj'}\ B_{\mu\mu'}\
         (w_{\lambda'\mu'})^*(v_{i'j'})^*      \cr
   &=\sum_{jj'} v_{ij} A_{jj'}X_{\lambda\lambda'} (v_{i'j'})^* \cr
\hbox{where}\qquad
 X_{\lambda\lambda'}
   &=\sum_{\mu\mu'}
         w_{\lambda\mu}\
         B_{\mu\mu'}\ (w_{\lambda'\mu'})^*     \cr
   &=(w(B\otimes\idty_\C)w^*)_{\lambda\lambda'}\ \in\C
\quad.\cr}$$

Suppose that, for some $B\in\M k\Cx$, and
$A_{jj'}=\delta_{jj'}$, the above matrix element contains a factor
$\delta_{ii'}$. We can rewrite this as
$\sum_jv_{ij}X_{\lambda\lambda'}(v_{i'j})^*=\delta_{ii'}\widetilde
X_{\lambda\lambda'}$, with $\widetilde X_{\lambda\lambda'}\in\C$.
In basis free formulation this reads
$v(\idty_d\otimes X_{\lambda\lambda'})v^*
  =\idty_d\otimes \widetilde X_{\lambda\lambda'}$, for all
$\lambda,\lambda'$. This condition can be considered for each pair
$\lambda\lambda'$ separately, and yields, in the special case of
$\SnU2$, that $X_{\lambda\lambda'}=\widetilde
B_{\lambda\lambda'}\idty_\C$ for some $\widetilde
B_{\lambda\lambda'}\in\Cx$ (see the Proposition below). But then, by
applying the counit $\counit$ to the definition of $X$, we find that
$$ \widetilde B_{\lambda\lambda'}
   =\counit\bigl(X_{\lambda\lambda'}\bigr)
    =\sum_{\mu\mu'}
         \counit\bigl(w_{\lambda\mu}\bigr)\
         B_{\mu\mu'}\ \counit\bigl((w_{\lambda'\mu'})^*\bigr)
    =B_{\lambda\lambda'}
\quad.$$
Hence  $X=(w(B\otimes\idty_\C)w^*)=B\otimes\idty_\C$, i.e.\ $B$ is
invariant under $w$.
\QED

The special property of $\SnU2$ used in the proof of (3) is isolated
in the following Proposition. It is clearly violated for ordinary
groups, for which (3) fails accordingly. In a sense it expresses the
property that $\SnU2$ is ``completely quantum''. In particular, it
shows that the center of $\SnU2$ consists only of multiples of the
identity.

\iproclaim/SUq/ Proposition.
Let $v$ be a non-trivial $d$-dimensional unitary representation
of $\SnU2$ with $d<\infty$, and let $X,\widetilde X\in\SnU2$ such that
$$ v(\idty_d\otimes X)v^*=\idty_d\otimes\widetilde X
\quad.$$
Then $X=\widetilde X$ is a multiple of the identity in $\SnU2$.
\eproclaim

\proof:
$v$ contains a non-trivial irreducible subrepresentation, hence
we may assume without loss of generality that $v$ is irreducible,
say, the irreducible representation of dimension $d=(2s+1)$, $s>0$.
Moreover, by applying the result to hermitian and skew-hermitian
parts, we can assume that $X$, and consequently $\widetilde X$,
is hermitian. By multiplying the equation from the right by $v$ we
get $v(\idty_d\otimes X)=(\idty_d\otimes\widetilde X)v$, or
$$ v_{ij}X=\widetilde Xv_{ij}
\quad,$$
for all $i,j$.

In order to make use of this condition we have to obtain information
about the matrix elements of the $(2s+1)$-dimensional, or
``spin-$s$''- representation of $\SnU2$. We use the standard
notation $\ket s,m>$, $m=-s,\ldots,s$ for the basis vectors of this
representation. We can realize it as that subrepresentation of the
$2s$-fold tensor product of the defining spin-$\half$ representation
$u$ with itself, which contains the
product vectors $\Psi_+=\ket\half,\half>^{\otimes 2s}$, and
$\Psi_-=\ket\half,-\half>^{\otimes 2s}$, and these vectors are
identified with $\ket s,\pm s>$, respectively.
Hence
$$ \bra s,s \abs{\,v\,} s,-s>
           =\bra\Psi_+,u^{\otimes 2s}\Psi_->
           =\bra \half,\half \abs{\,u\,} \half,-\half>^{2s}
           =\gamma^{2s}
\quad,$$
and, similarly,
$$\eqalign{
      \bra s,-s \abs{\,v\,} s,s>
           &=(-\nu)^{2s}\gamma^{*\,2s} \cr
       \bra s,s \abs v s,-s+1>
           &=\hbox{const}\times \alpha^*\gamma^{2s-1}
\quad.}$$

Now let $X$ and $\widetilde X$ be as in the Proposition, and
hermitian. Then $X\gamma^{2s}=\gamma^{2s}\widetilde X$, and
$X\gamma^{*\,2s}=\gamma^{*\,2s}\widetilde X$, which implies
$\gamma^{2s}X=\widetilde X\gamma^{2s}$. Hence $X$ commutes with
$(\gamma\gamma^*)^{2s}$. Similarly, we conclude that $X$ commutes
with even powers of $\alpha^*\gamma^{2s-1}$.

The irreducible representations of the C*-algebra of $\SnU2$ are
well-known \break \cite{WICK,Woronowicz}. In particular, one obtains a
faithful family of representations $\pi_{\zeta}$, parametrized by a
phase $\zeta\in\Cx$, by starting from a cyclic vector
$\Omega\in\H_\zeta$ with
$\pi_{\zeta}(\alpha)\Omega=0$, and setting
$$ \pi_{\zeta}(\gamma)\ \pi_{\zeta}(\alpha^{*n})\Omega
       = \zeta\,\nu^n\ \pi_{\zeta}(\alpha^{*n})\Omega
\quad.$$
The mutually orthogonal vectors $\pi_{\zeta}(\alpha^{*n})\Omega$,
$n\in\Nl$ span the representation space $\H_\zeta$.
Since the spectrum of $\pi_\zeta(\gamma\gamma^*)$ is simple, and
$\pi_\zeta(X)$ commutes
with this operator, $\pi_\zeta(X)$ is determined by its eigenvalues
$\xi_n$ via
$$ \pi_\zeta(X\alpha^{*n})\Omega
     =\xi_n\ \pi_\zeta(\alpha^{*n})\Omega
\quad.$$
Since $X$ commutes with even powers of $\alpha^*\gamma^{2s-1}$, it
commutes with $\alpha^{*2}$, and hence $\xi_{n+2}=\xi_n$. Hence
$\pi_\zeta(X)$ is a linear combination of the identity and the
unitary operator $U$ determined by
$$ U\pi_\zeta(\alpha^{*n})\Omega
    =(-1)^n\ \pi_\zeta(\alpha^{*n})\Omega
\quad.$$
The coefficient of $U$ must be zero, because $U$ is not in the
C*-algebra generated by $\pi_\zeta(\alpha)$ and $\pi_\zeta(\gamma)$.
To see this, consider the images of $U$, $\pi_\zeta(\alpha)$, and
$\pi_\zeta(\gamma)$ in the Calkin algebra, i.e.\ the quotient of
$\B(\H)$ by the algebra of compact operators. There the compact
operator $\pi_\zeta(\gamma)$ becomes zero, so $\pi_\zeta(\alpha)$
becomes unitary, and the algebra generated by these two becomes
abelian. On the other hand,
$U\pi_\zeta(\alpha)=-\pi_\zeta(\alpha) U$, hence the image of $U$
cannot be in this abelian algebra.

Hence $\pi_\zeta(X)=f(\zeta)\idty$, and we have to show that $f$ is
constant. Different representations are connected via
$\pi_\zeta\circ\Phi_t=\pi_{\zeta+t}$, where  $\Phi_t$ are the
automorphisms defined by $\Phi_t(\alpha)=\alpha$, and
$\Phi_t(\gamma)=\exp(it)\gamma$.
The $n$\th Fourier coefficient of $f$ is determined by the element
$X_n=(2\pi)^{-1}\int dt\ \exp(-int)\Phi_t(X)\in\SnU2$. Recall that
$X$ may be approximated in norm by polynomials $X^\epsilon$ in
$\alpha$, $\gamma$, and their adjoints. Using the relations \eq(Snu2)
we can
bring every approximating polynomial into a form in which no
monomial contains both $\gamma$ and $\gamma^*$. Then the above
integral picks out precisely those terms from any polynomial
containing $n$ factors $\gamma$ (or $-n$ factors $\gamma^*$). Let
$X_n^\epsilon$ denote the sum of these terms. Since
$\pi_\zeta(\gamma)$ is a compact operator it follows that
$\pi_\zeta(X_n^\epsilon)$ is compact for $n\neq0$, and, by norm
approximation, so is $\pi_\zeta(X_n)$. On the other hand,
$\pi_\zeta(X_n)$  is a multiple of the identity, and hence must be
zero for $n\neq0$. It follows that all Fourier coefficients of $f$
except the $0$\th vanish, and so $f$ is constant.
\QED

We now come to the discussion of the consequences of \Thm/2pts/ and of the
definition of action. Items (1) and (2) of \Thm/2pts/ can both be used to
define structures on infinite systems. Let us fix the algebra $\A=\M d\Cx$
as the observable algebra at each site of a lattice system, and a unitary
representation $v\in\M d\C\equiv\A\otimes\C$ of the quantum group \qg. The
observable algebra associated with a finite subset $\Lambda$ of the lattice
under consideration is then $\A^\Lambda=\bigotimes_{i\in\Lambda}\A\up i$,
where $\A\up i$ is an isomorphic copy of $\A$. By $\idty^\Lambda$ we denote
the identity element in this algebra. If $\Lambda=\Lambda_1\cup\Lambda_2$
is the disjoint union of two subregions, we have a canonical isomorphism
$\A^{\Lambda}\cong\A^{\Lambda_1}\otimes\A^{\Lambda_2}$. For
$\Lambda_1\subset\Lambda_2$ we have the inclusion
$\A^{\Lambda_1}\subset\A^{\Lambda_2}$, where the inclusion map is $A\mapsto
A\otimes\idty^{\Lambda_2\setminus\Lambda_1}$. For an infinite set $\Lambda$
we can therefore consider the union of all algebras $\A^{\Lambda_f}$ for
finite $\Lambda_f\subset\Lambda$. This algebra carries a natural C*-norm,
and we will denote by $\A^\Lambda$ the {\it C*-inductive limit} of the
$\A^{\Lambda_f}$, i.e.\ completion of the union in this norm. As a special
case, we obtain the observable algebra, also called the {\it quasi-local
algebra} of the infinite lattice system \cite{BraRo}, by taking
$\Lambda$ as the whole lattice.

In order to define a quantum group action on $\A^\Lambda$ we begin with the
case of finite $\Lambda$. The unitary representation
$v\tee v\cdots \tee v$ ($n$
times) is easily seen to be independent of the bracketing of the
$\tee$-products, hence we can define the action
$\alpha_{(v\tee\cdots \tee v)}$
on $\A^{\bracks{1,n}}$. Note that, in contrast to the case of ordinary
groups, the ordering of sites in this product is essential, since it
fixes the ordering of factors in $\C$. This means that actions of
quantum groups can only be defined on one-dimensional lattice systems.
Analogously, in quantum field theory, the typical applications of
quantum groups are to systems in one space and one time dimension.

In order to define an action on a quasi-local algebra, we have to
use the inductive limit process. Thus we would like to define
$\alpha(A)=\alpha_{(v\tee\cdots \tee v)}(A)$, whenever
$A\in\A^{\bracks{m+1,m+n}}$, i.e.\ $A$ is in an algebra belonging to
$n$ consecutive sites. This preliminary definition has to be checked
for consistency with the inclusion maps
$A\mapsto A\otimes\idty^{\Lambda_2\setminus\Lambda_1}$, i.e.\ we
have to verify that we obtain the same result if we consider $A$ as
an element of a larger algebra $\A^{\bracks{m+1-\ell,m+n+r}}$ with
$\ell,r\geq0$. This is precisely the function of \Thm/2pts/(1): it
shows that consistency holds for arbitrary $r$ and $\ell=0$. On the
other hand, \Thm/2pts/(3) shows that for a proper quantum group
consistency fails on the left, i.e. for $\ell>0$. The best we can do
is therefore to define an action $\alpha^\Nl_v$ on the half-infinite chain
$\A^\Nl$, setting
$$ \alpha^\Nl_v(A)= \alpha_{(v\tee\cdots \tee v)}(A)
\qquad\hbox{for}\quad A\in\A^{\bracks{1,n}}
\quad.\deqno(actNl)$$

There is a very elegant way of constructing this action
\cite{Watatani}, which also underlines the special role of the half
chain in this context: the algebra $\A^\Nl$ with $\A=\M d\Cx$ can be
considered as the gauge invariant part of the Cuntz algebra $\Cuntz_d$
\cite{Cuntz}. This is the algebra generated by $d$ Hilbert space
operators $S_i$, $i=1,\ldots,d$ satisfying the relations
$$\eqalign{             S_i^*S_j&=\delta_{ij}\idty
\quad,\qquad\hbox{for}\quad i,j,=1,\ldots,d \cr
           \sum_{i=1}^d S_iS_i^*&=\idty\quad.\cr}
\deqno(Cuntzrel)$$
The algebra generated by such operators is independent of the
realization, in the sense that for any C*-algebra $\A$, and any
elements $\widetilde S_i\in\A$ satisfying the same relations, there is
a unique injective C*-homomorphism $\Phi:\Cuntz_d\to\A$ such that
$\Phi(S_i)=\widetilde S_i$. In particular, there is a one-parameter
automorphism group $\gamma_t$ on $\Cuntz_d$ such that
$\gamma_t(S_j)=e^{it}S_j$. The fixed point algebra of this action is
called the {\it gauge invariant part} of $\Cuntz_d$. It is
canonically isomorphic to the half chain algebra $\A^\Nl$, because the
operators
$$ S_{i_1}S_{i_2}\cdots S_{i_n}S_{i_n}^*S_{i_{n-1}}^*\cdots S_{i_1}^*
   \qquad\hbox{for $i_\nu=1,\ldots,d$,}
$$
satisfy precisely the algebraic relations of the matrix units in
$\A^{\bracks{1,n}}$. Moreover, these matrix units are compatible with
tensoring of identity operators on the right, because in the above
expression the sum over $i_n$ leaves the corresponding expression for
$n'=n-1$. The idea of \cite{Watatani} for obtaining an action of a
quantum group on $\A^\Nl$ is to define an action $\alpha^{\Cuntz_d}$ on
$\Cuntz_d$ instead, which restricts to $\A^\Nl$, because the action
commutes with $\gamma_t$. Given a unitary representation $v\in\M d\C$
of the quantum group \qg, they define
$$ \alpha^{\Cuntz_d}(S_i)=\sum_{j=1}^d S_j\otimes v_{ji}
\quad\in\Cuntz_d\otimes\C
\quad.\deqno(Watata)$$
The existence of a unique injective C*-homomorphism
$\alpha^{\Cuntz_d}$ with this property follows at once from the
universal property of $\Cuntz_d$, by verifying that the right hand
side satisfies the relations \eq(Cuntzrel). By considering the
action on matrix units it becomes clear that this action is the same
as the one constructed above. An extension of this construction to
the doubly infinite chain is impossible, since the identification of
the matrix units, and hence of $\A^\Nl$ as the gauge invariant part
of $\Cuntz_d$ breaks down.

We argued at the beginning of this chapter that for the definition
of tensor products it is more natural to consider extended actions
$\extdact:\A\otimes\C\to\A\otimes\C$, rather than simple actions
$\alpha:\A\to\A\otimes\C$. The drawback of this approach to the
tensor product of actions is again in the issue of locality: with
the simpler notion based on tensoring of unitary representations, we
had locality problems at the left end of the chain. With the
approach based on extended actions, we get problems right and left.
In particular, the action on the half chain fails to meet the higher
standards for extended actions.

\iproclaim/P.non-ext/ Proposition.
\item{(1)}
Let $\alpha^\Nl_v$ be the action on the half chain associated with
the irreducible representation $v$ of $\SnU2$. Then the relative
commutant of $\alpha^\Nl_v(\A^\Nl)\subset\A^\Nl\otimes\C$ consists
only of multiples of the identity.
\item{(2)}
The action $\alpha^\Nl_v$  is not extendible.
\eproclaim

\proof:
Suppose that $X\in\A^\Nl\otimes\C$ commutes with
$\alpha^\Nl_v(\A^\Nl)$. We have to show that $X=x\idty$. Let
$\omega$ be a state on $\A$, and let
$\E_N:\A^\Nl\otimes\C\to\A^\Nl\otimes\C$ be the conditional
expectation defined by
$$ \E_N(A_N\otimes A'\otimes C)
      =\omega^{\otimes\infty}(A')\
       A_N\otimes\idty\otimes C
\quad,$$
where the tensor product refers to the decomposition
$\A^\Nl\otimes\C=\A^{\bracks{1,N}}\otimes\A^{\lbrack N,\infty)}\otimes\C$,
and $\omega^{\otimes\infty}$ denotes the infinite product state.
Then, since $X\in\A^\Nl\otimes\C$, the sequence $X_N=\E_N(X)$
converges in norm to $X$. Moreover, $X$, and hence $X_N$ commutes
with $v^{\tee N}(A_N\otimes\idty\otimes\idty_\C)v^{\tee N\,*}$, for
$A_N \in\A^{\bracks{1,N}}$. Since $v^{\tee N}$ is unitary, this
means that $v^{\tee N\,*}X_N v^{\tee N\,*}$ commutes with all $A_N$.
Hence this element must be in
$\idty_N\otimes\A^{\lbrack N,\infty)}\otimes\C$.
By definition of the conditional expectation, it is also in
$\A^{\bracks{1,N}}\otimes\idty\otimes\C$. Hence there is some
$C_N\in\C$ such that
$$ v^{\tee N\,*}X_N v^{\tee N\,*} =\idty_N\otimes\idty\otimes C_N
\quad.$$
Using the relation $X_N=\E_N(X_{N+1})$, we find the formula
connecting the different $C_N\in\C$:
$$ C_N= \omega\otimes\id \bigl(v(\idty_\A\otimes C_{N+1})v^*\bigr)
\quad.\deqno*()$$

We will show, in the special case of $\SnU2$, that this implies
$C_N=c\idty_\C$ for all $N$. Then $X_N=c\idty$, and
$X=\lim_NX_N=c\idty$. The non-extendibility (2) of $\alpha_v^\Nl$
follows from statement (1): for if
$\widehat\alpha\in\Aut(\A^\Nl\otimes\C)$ is an automorphism
extending $\alpha_v^\Nl$, every element of the form
$\widehat\alpha(\idty\otimes C)$ is in the commutant of
$\widehat\alpha(\A\otimes\idty_\C)=\alpha_v^\Nl(\A)$. Hence by the
determination of the commutant, there must be a linear functional
$\eta:\C\to\Cx$ such that
$$\widehat\alpha(A\otimes C)
    =\alpha_v^\Nl(A)\eta(C)
\quad,$$
which clearly contradicts $\alpha$ being an automorphism. Even if we
do not insist on the invertibility of $\widehat\alpha$, and allow
more general homomorphisms satisfying \eq(actionp), we find from
\eq(actionp) that $\eta$ must be a one-dimensional representation of
the quantum group, i.e.\ typically $\eta=\counit$. This choice once
again contradicts \eq(actionp).

It remains to prove that equation \eq$\ast$() implies that all $C_N$
are multiples of the identity, assuming that $v$ is the spin-$s$
representation of $\SnU2$. We are free to choose the state $\omega$
for convenience, and we will take $\omega$ as the pure state with
highest $3$-component of the spin. Then
$$\eqalign{
  \Phi(C)&=\omega\otimes\id \bigl(v(\idty_\A\otimes C)v^*\bigr) \cr
         &=\sum_{m=-s}^{+s}v_{s,m}Cv_{s,m}{}^*                  \cr
         &=\sum_{m=-s}^{+s} \lambda_m (\alpha^{s+m})^*\,
              (\gamma^{s-m})^*\ C \gamma^{s-m}\,\alpha^{s+m}
\quad,}$$
where the $\lambda_m$ are strictly positive constants. We will
evaluate condition \eq$\ast$(), i.e.\ $C_N=\Phi(C_{N+1})$ in the
faithful family of representations $\pi_\zeta$ used in the proof of
\Prp/SUq/. Thus, denoting an orthonormal basis of the representation
space by $\ket n>$, with $n=0,1,\ldots$, and $\Omega=\ket0>$, we
have
$$\eqalign{
   \pi_\zeta(\alpha)\ket n>&= \sqrt{1-\nu^{2n}}\, \ket n-1>  \cr
   \pi_\zeta(\alpha^*)\ket n>&= \sqrt{1-\nu^{2n+2}}\, \ket n+1>  \cr
   \pi_\zeta(\gamma)\ket n>&= \zeta\,\nu^n\, \ket n>
\quad.}$$
It follows that the matrix element $\bra n\abs{\Phi(C)}m>$ depends
only on the matrix elements $\bra n'\abs{C}m'>$ with $n'\leq n$,
$m'\leq m$, and $n-m=n'-m'$. The iteration of $\Phi$ thus breaks
down into a family of finite dimensional iterations of a triangular
matrix with positive entries. Each of these operators is
contractive, and has a unique fixed point, which is zero for $n\neq
m$, and a vector with constant entries for $n=m$. Now for each $N$,
$C_N=\Phi^{M}(C_{N+M})$, with
$\abs{\bra n\abs{C_{N+M}}m>}\leq\norm{X}$, for all $n,m$.
Since $M$ can be chosen arbitrarily large, each matrix element of
$C_N$ must be arbitrarily close to a fixed point. Hence $C_N$ is a
multiple of the identity.
\QED

\vfill\eject

\bgsection 4. Invariance of observables and states

In the previous section we studied the difficulties in extending the
notion of group action to an infinite chain. The basic problem was
that the family of actions, defined for each finite segment, are not
compatible with the identifications used for the C*-inductive limit
by which the algebra of the whole chain is defined. In this section we
will see that some derived structures, defined on finite segments in
terms of the quantum group action, may nevertheless be compatible with
the inductive limit.

The most important case in point is the notion of invariant elements
under the action: by \Thm/2pts/ the locality property
$\alpha\bigl(\A^\Lambda\bigr)\subset\A^\Lambda\ot\C$ does hold for
the quantum group invariant elements. This allows us to make the
following definition:

\iproclaim/invEl/ Definition.
Let $\A=\M d\Cx$, and $v$ a $d$-dimensional unitary representation of
a quantum group \qg. Let $m\in\Ir$, and $n\in\Nl$, and let
$A\in\A^{\bracks{m+1,m+n}}$.
Then $A$ is called $\ad(v)$-invariant, if $A$ is an intertwiner for
$v\tee\cdots \tee v$, or equivalently, if
$\ad(v\tee\cdots \tee v)(A)=A\otimes\idty_\C$.
\eproclaim

The point is that this definition is independent of the local
algebra in which we consider $A$, i.e.\ the $\ad(v)$-invariance of
$A$ implies the invariance of $\idty_\A\otimes A$, and
$A\otimes\idty_\A$, which are considered to be ``the same element''
in the quasi-local algebra. For ordinary groups, the notion of
$\ad(v)$-invariance is equivalent to the invariance of $A$ under the
action $\ad(v^{\otimes\infty})$ on the whole lattice system.
However, as we have seen, this action on the whole chain $\A^\Ir$ is
not well-defined in the quantum group case.
Thus the invariance in \Def/invEl/ is {\it not} the invariance with
respect to a fixed action of the quantum group.

The structure of the algebra of invariant elements on a finite chain
is determined essentially by the reduction theory of the tensor
product representations $v^{\tee n}$ into irreducibles ones. In the case
of $\SnU2$ these decompositions are isomorphic to those for the
classical group $\SU2$. Therefore the inductive limits of the
algebras of invariant elements are also isomorphic in the deformed
and undeformed case.

The second notion we are interested in is that of invariant states.
We saw we cannot define the quantum group invariance of a translation
invariant state as invariance under an action, simply because such
actions don't exist. However, just as in the case of invariant
observables we may define this property by considering only a finite
subchain at a time.

\iproclaim/hered/ Definition.
Let $\A=\M d\Cx$, and $v$ a $d$-dimensional unitary representation of
a quantum group \qg. Then a state $\omega$ on $\A^\Ir$ is called {\bf
hereditarily invariant}, if, for all $n\leq m\in\Ir$, the restriction
$\omega\rstr\A^{\bracks{n,m}}$ is invariant with respect to
$\alpha_{v^{\tee m-n+1}}$.
\eproclaim

It is easy to see that if $\omega$ is invariant for
$\alpha_{v\tee w}$, its restriction to the first tensor factor is
$\alpha_v$-invariant. From examples one can see that its restriction
to the second factor is not necessarily $\alpha_w$-invariant. On the
other hand, hereditarily invariant states do exist: if
$\omega=\omega_1\otimes\omega_2$ with $\omega_1$
$\alpha_v$-invariant, and $\omega_2$ $\alpha_w$-invariant, then
$\omega$ turns out to be $\alpha_{v\tee w}$-invariant. Hence the
infinite product state formed with an $\alpha_v$-invariant state at
each site of an infinite chain is hereditarily invariant. Since the
reduction theory of tensor products, and hence the the decomposition
rules for invariant states are the same for $\SnU2$ as for
$\SU2$ one might expect that, as in the classical case, there may be
many hereditarily invariant states. However, once more the
$\nu$-deformation spoils this expectation. In explicit
computations (spin-1/2 chain up to length $6$, spin-1 chain up to
length 3, and some tensor products of other irreducible
representations) we found that {\it only} the product state is
hereditarily invariant.
We were not able, however, to decide the following statement:

\iproclaim/conject/ Conjecture.
Let $v$ be an irreducible unitary representation of $\SnU2$. Then the
only hereditarily invariant state of $\A^\Ir$ is the product state
$\omega_1^{\otimes\infty}$ formed with the $\alpha_v$-invariant state
$\omega_1$ at each single site.
\eproclaim

\vfill\eject
\bgsection 5. Quasi-local actions

The arguments of the previous paragraphs show that local actions of
quantum groups cannot be obtained using the recipes familiar from
classical groups. We will now show that these difficulties are
inherent in the quantum group concept, i.e.\ local actions do not
exist on general grounds. Of course, there is always the trivial
action of a quantum group, which is obviously local. For a classical
group we would exclude such trivialities by assuming the action to
be {\it faithful}, i.e.\ that the only group element $g$ with
$\alpha_g(A)=A$ for all $A$ is the identity. The following
Proposition shows how to say this for the action of a quantum group.
The condition of nuclearity is automatically satisfied for the
quasi-local algebra of a spin chain. It implies that the minimal and
maximal C*-tensor products \cite{Takesaki} of $\A$ with any other
C*-algebra coincide. In particular, the minimal tensor product
$\A\otimesmin\C$ we have been using in the definition of actions is
the same as the maximal one.

\iproclaim/quot/ Proposition.
Let $\alpha:\A\to\A\otimes\C$ be an action of a quantum group \qg\
on a nuclear C*-algebra $\A$. Then there is a smallest C*-subalgebra
$\Csub\subset\C$ such that $\alpha(\A)\subset\A\otimes\Csub$.
$\Csub$ is closed under the coproduct in the sense that
$\copr(\Csub)\subset\Csub\otimes\Csub$, and is hence a quantum group
in its own right. In the classical case $\C=\C(G)$ it is the algebra
of functions on the quotient of $G$ by the subgroup of all $h$ such
that $\alpha_h(A)=A$ for all $A\in\A$.
\eproclaim

For the proof we need a fact about nuclear C*-algebras, which we
summarize in a Lemma. When $\rho\in\A^*$ is a linear functional,
denote by ``$\rho\otimes\id_\C$'' the continuous linear extension of
$(\rho\otimes\id_\C)(A\otimes C)=\rho(A)C$. For a state $\rho$ this
is the conditional expectation onto the second factor.

\iproclaim/nucLem/ Lemma.
Let $\A$ be a nuclear C*-algebra, $\C$ another C*-algebra, and
$\D\subset\A\otimes\C$ a closed subspace. Let $\Csub\subset\C$ be
the closed subspace generated by all elements of the form
$(\rho\otimes\id_\C)(D)$, where $\rho\in\A^*$, and $D\in\D$.
Then $\Csub$ is the smallest closed subspace with the property
$$ \D\subset \overline{\A\otimes\Csub}^{\norm{\cdot}}
\quad.$$
\eproclaim

\proof{ of the Lemma:}
We first show that $\Csub$ has the stated property. Since $\D$ and
$\overline{\A\otimes\Csub}^{\norm{\cdot}}$ are norm closed
subspaces, the inclusion given is equivalent to
$$ (\A\otimes\Csub)^{\perp}\subset \D^{\perp}
\quad,\deqno*()$$
where $\D^{\perp}$ denotes the space of functionals in
$(\A\otimes\C)^*$ annihilating a subspace $\D$. We now show that,
for any finite rank operator $F:\A^*\to\A^*$, we have
$$ (F\otimes\id_\C)(\A\otimes\Csub)^{\perp}\subset\D^{\perp}
\quad.\deqno**()$$
Since $F$ is of finite rank, it is of the form
$$ \bra F\omega,A>_{\A\A^*}
       =\sum_{i=1}^N \bra\omega,X_i>_{\A^*\A^{**}}\
                     \bra\omega_i,A>_{\A^*\A}
\quad,$$
where the brackets denote the canonical bilinear forms of the
pairings indicated, and $X_i\in\A^{**}$, $\omega_i\in\A^*$.
This can also be expressed conveniently as a map $\widetilde F$ on
density matrices $D_\omega$
(defined by $\bra\omega,A>=\tr D_\omega\pi(A)$) in the universal
representation $\pi$ of $\A$:
$$ \widetilde F(D_\omega)
       =\sum_{i=1}^N \tr(D_\omega X_i)\ D_{\omega_i}
\quad,$$
where $X_i$ is now considered as an element of the weak closure of
$\pi(\A)$. The operator $F\otimes\id_\C$ can be expressed similarly
by its action on density matrices in the representation
$\pi\otimes\pi_\C$, where $\pi_\C$ is any faithful representation of
$\C$. One gets
$$ \bra(F\otimes\id_\C)\Omega,\ Y>
      :=\sum_{i=1}^N \tr\bigl(D_\Omega\, X_i\otimes
                             \pi_\C((\omega_i\otimes\id_\C)(Y))\bigr)
\quad.$$
Now, if $\Omega\in(\A\otimes\Csub)^{\perp}$, and $Y\in\D$, we have
$\tr\bigl(D_\Omega\, X_i\otimes
          \pi_\C(\omega_i\otimes\id_\C(Y))\bigr)=0$ for $X_i\in\pi(\A)$,
and this extends to $X_i\in\A^{**}$, identified with the weak
closure of $\pi(\A)$. Hence $(F\otimes\id_\C)\Omega\in\D^{\perp}$,
which proves \eq$\ast\ast$().

By nuclearity, the identity on $\A^*$ is the simple weak*-limit
of a net $F_{\alpha}$ of completely positive normalized, finite rank
operators \cite{CEffros}. Since the $F_\alpha$ are uniformly bounded
this implies that the identity on $(\A\otimes\C)^*$ is the limit of
the net $(F_\alpha\otimes\id_\C)$. That is, for
$\Omega\in(\A\otimes\C)^*$ we have
$$ w^*{-}\lim_\alpha (F_\alpha\otimes\id_\C)\Omega=\Omega
\quad.$$
Hence, for $\Omega\in(\A\otimes\Csub)^{\perp}$ the preceeding
paragraph implies $\Omega\in\D^{\perp}$, proving \eq$\ast$().

It remains to be shown that
$\Csub$ is the smallest subspace with the stated property. Suppose
that $\Csubb$ also satisfies
$\D\subset \overline{\A\otimes\Csubb}^{\norm{\cdot}}$, and let
$D\in\D$. Then we can write $D$ as the norm limit of elements
$$ D_\alpha=\sum_{i=1}^{N_\alpha} A_\alpha\otimes C_\alpha
\quad,\quad\hbox{with $C_\alpha\in\Csubb$.}$$
Thus, for any state $\rho\in\A^*$, we get
$(\rho\otimes\id_\C)(D_\alpha)=\sum_i\rho(A_\alpha)C_\alpha\in\Csubb$,
and, since $(\rho\otimes\id_\C)$ is a contraction:
$(\rho\otimes\id_\C)(D)\in\Csubb$. Consequently,
$\Csub\subset\Csubb$.
\QED

\proof{ of the Proposition:}
It is clear that the Lemma remains valid, if we demand $\Csub$ to be
a C*-subalgebra rather than a closed subspace. Hence we can take
$\Csub$ as the C*-subalgebra of $\C$ generated by all elements of
the form $(\rho\otimes\id_\C)(\alpha(A))\in\C$. It remains to be
shown that
$\copr(\rho\otimes\id_\C)\alpha(A)\subset\Csub\otimes\Csub$. By the
action property of $\alpha$ we have
$$\eqalign{
  \copr(\rho\otimes\id_\C)\alpha(A)
   &=\copr(\rho\otimes\id)\alpha(A)
    =(\rho\otimes\id_\C\otimes\id_\C)(\id_\A\otimes\copr)\alpha(A) \cr
   &=(\rho\otimes\id_\C\otimes\id_\C)(\alpha\otimes\id_\C)\alpha(A)
    =((\rho\otimes\id_\C)\alpha\otimes\id_\C)\alpha(A) \cr
   &\in((\rho\otimes\id_\C)\alpha\otimes\id_\C)\A\otimes\Csub
    \subset\Csub\otimes\Csub
\quad.}$$
Since the $(\rho\otimes\id_\C)\alpha(A)$ generate $\Csub$, and
$\copr$ is a *-homomorphism, we find
$\copr(\Csub)\subset\Csub\otimes\Csub$.

When $\C=\C(G)$ is abelian, any C*-subalgebra $\Csub$ is uniquely
characterized by the equivalence relation $g\approx g'$ defined by
$f(g)=f(g')$ for all $f\in\C(G)$. In the present case this becomes
$\rho(\alpha_g(A))=\rho(\alpha_{g'}(A))$ for all $A$ and all
$\rho$, i.e.\ $\alpha_g=\alpha_{g'}$. Thus $\Csub$ is the algebra
of functions on the quotient of $G$ by the subgroup acting trivially
on $\A$.
\QED

Consider now an action on a C*-algebra $\A$, containing two ``local''
subalgebras $\A_1$ and $\A_2$, by which we only mean in the present
context that they commute elementwise. The action is called {\it
strictly local}, if $\alpha(\A_i)\subset\A_i\otimes\C$, which is the
notion considered in the previous section. The action is
called {\it local}, if $\alpha(\A_i)\subset\A_i^\odot\otimes\C$, where
$\A_i^\odot\supset\A_i$, $i=1,2$, are two algebras which still commute
elementwise. The typical situation we have in mind here is that the
$\A_i$ are the algebras belonging to two disjoint finite regions in
the lattice of a spin system, and the $\A_i^\odot$ belong to two
larger, but still disjoint regions. We can consider a still weaker
condition, which does not require $\alpha(A)$ to be localized in any
finite region, but allows a weak delocalized tail. We call the action
{\it quasi-local} if, for any localized $A$,
$\alpha(A)\in\Aql\otimes\C$, where $\Aql$ denotes the quasi-local
algebra of the spin system, i.e.\ the C*-inductive limit of the local
algebras. Then by the norm continuity of $\alpha$, we have
$\alpha(\Aql)\subset\Aql\otimes\C$. Thus quasi-locality of an action
of a quantum group just means that it can be considered as an action
on the quasi-local C*-algebra.

\iproclaim/qloc/ Theorem.
Let $\Aql$ be the quasi-local observables of a spin system on an
infinite (not half-infinite)
lattice, and let $\alpha:\Aql\to\Aql\otimes\C$ be the action of
a quantum group \qg\ on $\Aql$. Assume that $\alpha$ is faithful in
the sense that $\alpha(\Aql)\subset\Aql\otimes\Csub$ holds for no proper
C*-subalgebra $\Csub\subset\C$, and that
$\alpha\tau_x=(\tau_x\otimes\id_\C)\alpha$, for all $x$, where
$\tau_x$ denotes the automorphism of $\Aql$ of translation by the
lattice vector $x$.
Then the C*-algebra $\C$ is abelian.
\eproclaim

\proof:
For a continuous linear functional $\rho$ on $\Aql$, and $A\in\Aql$,
consider $(\rho\otimes\id_\C)\alpha(A)\in\C$ as in the proof of the
above Lemma. By assumption, elements of this form generate $\C$.
Therefore, we only have to show that  $(\rho\otimes\id_\C)\alpha(A)$
and $(\rho'\otimes\id_\C)\alpha(A')$ commute for all $A,A'\in\Aql$,
and $\rho,\rho'\in\Aql^*$.
Now, for every $\epsilon>0$, we can find expressions
$\alpha(A)=\sum_{i=1}^nA_i\otimes C_i+\Rest(\epsilon)$, and
$\alpha(A')=\sum_{i=1}^{n'}A'_i\otimes C'_i+\Rest(\epsilon)$,
where $A_i,A'_i\in\Aql$, $C_i,C'_i\in\C$, and here and in the sequel
$\Rest(\epsilon)$ stands for any rest which is bounded in norm by
$\epsilon$. In these expressions we may take the $A_i$ and $A'_i$ to
be localized in a finite subset $\Lambda$ of the lattice. Now let $x$
be a translation such that $\Lambda\cap(\Lambda+x)=\emptyset$, and
$\norm{\bracks{A,\tau_xA'}}\leq\epsilon$.
Then
$$\eqalign{
  \bracks{(\rho\otimes\id_\C)\alpha(A),
          (\rho'\otimes\id_\C)\alpha(A')}=\hskip-90pt&\cr
   &=\sum_{ij}\rho(A_i)\rho'(A'_j) \bracks{C_i,C'_j}
      +\Rest\bigl(2\epsilon(\norm{A}+\norm{A'}+3\epsilon)\bigr) \cr
   &=\sum_{ij}\widetilde\rho\bigl(A_i\otimes\tau_x(A'_j)\bigr)
       \bracks{C_i,C'_j}
      +\Rest\bigl(2\epsilon(\norm{A}+\norm{A'}+3\epsilon)\bigr) \cr
   &=(\widetilde\rho\otimes\id_\C)(\bracks{
          \alpha(A),\ (\tau_x\otimes\id_\C)\alpha(A')} )
      +\Rest\bigl(4\epsilon(\norm{A}+\norm{A'}+3\epsilon)\bigr) \cr
   &=(\widetilde\rho\otimes\id_\C)
          \alpha(\bracks{A, \tau_x A')} )
      +\Rest\bigl(4\epsilon(\norm{A}+\norm{A'}+3\epsilon)\bigr)
\quad,}$$
where $\widetilde\rho$ is a state on $\Aql$ which coincides with
$\rho$ in $\Lambda$ and with $\rho'\circ\tau_{-x}$ in
$(\Lambda+x)$.
Hence
$\norm{\bracks{(\rho\otimes\id_\C)\alpha(A),
               (\rho'\otimes\id_\C)\alpha(A')}}
  \leq \epsilon+4\epsilon(\norm{A}+\norm{A'}+3\epsilon)$, for any
$\epsilon$.
\QED

\vfill\eject
\bgsection 6. Quasi-local actions in a representation.

The aim of this section is to show the impossibility of constructing
in the GNS space of a translation invariant state, genuine quantum
group representations, commuting with the shift and sufficiently
local.

Let $\om$ be a translation invariant state on $\chain\A$ and
$(\H,\pi,\Om)$ the corresponding GNS space,
representation and cyclic vector. The translation automorphism $\tau$ is
implemented by the unitary shift $S$ on $\H$:
$$\pi(\tau(X)) = S\pi(X)S^* \quad \hbox{with}\quad
S\pi(X)\Om = \pi(\tau(X))\Om, \quad X\in\chain\A
\quad.\deqno(shift)$$
Let $U$ be a unitary representation of a quantum group \qg\ on $\H$.
For $\phi,\psi\in\H$, $A,C\in\B(\H)$ and $B,D\in\C$ we put
$$\bra A\ot B\phi,C\ot D\psi>
   =\bra A\phi,C\psi > B^*D
\quad,\deqno()$$
and extend this bilinearly to $\B(\H)\otimesmin\C$. This is possible
because we have for each representation $\tilde\pi$ of $\C$ on a
Hilbert space $\K$ and for all choices of $\phi,\psi\in\H$,
$\zeta,\eta\in\K$, $A_i,C_i\in\B(\H)$ and $B_i,D_i\in\C$,
$i=1,2,\ldots n$, $n=1,2,\ldots$
$$\eqalign{
 \bigl|\bra{\zeta,
     \sum_{k,\ell} \bra A_k\phi,C_\ell\psi>\
        \tilde\pi(B_k^*D_\ell) }\eta> \bigr|
\hskip-80pt& \cr
&= \bigl|\bra{
      \bigl(\sum_k    A_k   \ot\tilde\pi(B_k)  \bigr) \phi\ot\zeta,
      \bigl(\sum_\ell C_\ell\ot\tilde\pi(D_\ell)\bigr) \psi\ot\eta}>
     \bigr|  \cr
&\leq  \|\sum_k A_k\ot B_k\|\,  \|\sum_\ell C_\ell\ot D_\ell\|\,
       \norm{\phi}\,\norm{\psi}\,\norm{\zeta}\,\norm{\eta}
\quad. }$$
By $\norm{A\phi}^2$ we denote $\bra{A\phi,A\phi}>$, $\phi\in\H$ and
$A\in \B(\H)\otimesmin\C$. $U$ is said to act {\it almost locally}
if for any $\phi\in\H$, $\sigma\in\C^*$, $A$ and $B\in\chain\A$
$$\lim_{n\to\infty}
          \sigma\Bigl(\norm{[\pi(\tau^n(A)),U\pi(B)U^*]\phi}^2\Bigr)
         = 0
\quad.\deqno(alocal)$$

\iproclaim/inrep/ Proposition.
Let $\om$ be a translation invariant, clustering state on
$\chain\A$, with GNS triplet $(\H,\pi,\Om)$. Let $U$ be a unitary
representation of a quantum group \qg\ on $\H$ which commutes with
the shift $S$ on $\H$, acts almost locally and leaves $\Om$
invariant. Suppose that that there is no proper C*-subalgebra $\C_0$
of $\C$ such that $U\in\B(\H)\ot\C_0$. Then $\C$ is abelian.
\eproclaim

\proof:
Denoting by $S$ the
unitary on $\H$ that implements the shift, we can express clustering
as:
$$\lim_{n\to\infty} \bra{\phi,S^n\psi}> = \bra{\phi,\Om}>
\bra{\Om,\psi}>.$$
Choose now $A,B,C,D\in\pi(\chain\A)$. Using the asymptotic abelianness of
$\chain\A$, $US=SU$, $U\Om=S\Om=\Om$ and the almost locality of the action
of $U$, we compute:
$$\eqalign{
 \bra{AS^nB \Om, UCS^nD \Om}>
    &=\bra{A(S^nB(S^*)^n) \Om, UCS^nD \Om}> \cr
    &=\bra{A \Om, (S^nB^*(S^*)^n)UCS^nD \Om}> + \order(1) \cr
    &=\bra{A \Om, US^nU^*B^*U(S^*)^nCS^nD \Om}> + \order(1) \cr
    &=\bra{A \Om, UCS^nU^*B^*UD \Om}> \cr
    &\qquad+ \bra{A \Om, US^n [U^*B^*U,(S^*)^nCS^n] D \Om}>
          + \order(1) \cr
    &=\bra{C^*U^*A \Om, S^nU^*B^*UD \Om}> + \order(1) \cr
    &=\bra{C^*U^*A \Om,\Om}> \bra{\Om, U^*B^*UD \Om}>
          +  \order(1) \cr
    &=\bra{A \Om,UC\Om}> \bra{B\Om, UD \Om}> + \order(1)
\quad. }$$
Exchanging the roles of $A$ and $B$ and also of $C$ and $D$, replacing $n$
by $-n$, and using the asymptotic abelianness of $\chain\A$ we conclude:
$$\eqalign{
\bra{A \Om,UC\Om}> \bra{B\Om, UD \Om}>
&=\lim_{n\to\infty} \bra{AS^nB \Om, UCS^nD \Om}>  \cr
&=\lim_{n\to-\infty} \bra{BS^nA \Om, UDS^nC \Om}>  \cr
&=\bra{B \Om,UD\Om}> \bra{A\Om, UC \Om}>.
}$$
But this implies precisely the statement of the proposition.
\QED

\vfill\eject
\bgsection 7. \cfc\ states

The basic construction of states on a half chain and a chain that we will
use in this section is a generalization of the so-called Valence Bond Solid
states \cite{AKLT}. It was first given in \cite{FCS} and is based on an
earlier proposal for the construction of quantum Markov states in
\cite{AFri}. Apart from quasi-free CAR-states, it is the only construction
that we know of for obtaining non-product pure translation invariant states
on a spin chain.

We will assume throughout this section that the single-site observable
algebra $\A$ is that of the complex $d\times d$ matrices $\M d\Cx={\cal
M}_d$. A state $\om$ of the left half chain $\A^-=\A^{\Ir\setminus\Nl}$ is
completely determined by giving the expectation values
$\om^{\bracks{-n,-1}}(A)$ of observables $A\in\A^{\bracks{-n,-1}}$,
$n=1,2,\cdots$. The prescription for $\om^{\bracks{-n,-1}}(A)$ must be
compatible with the obvious requirement that
$\om^{\bracks{-n-1,-1}}(\idty\ot A)= \om^{\bracks{-n,-1}}(A)$. If we are
furthermore able to give a construction such that also
$\om^{\bracks{-n-1,-1}}(A\ot\idty)= \om^{\bracks{-n,-1}}(A)$, then we have
in fact defined a translation invariant state on the entire chain by
putting $\om(A) = \om^{\bracks{-n,-1}}(A)$, $A\in\A^{\bracks{m-n,m-1}}$,
$m\in\Ir$, $n=1,2,\cdots$.

Let $\B$ be a $\ast$-subalgebra of the $k\times k$ matrices ${\cal M}_k$,
containing the identity $\idty$ of ${\cal M}_k$ and let $\E$ be a unity
preserving, \cp\ map from $\B\ot\A$ to $\B$. Tensoring $\E$ with suitable
identity maps on factors $\A$, we can iterate $\E$ to obtain, for
$n\in\Nl$, unity preserving, \cp\ maps $\E^{(n)}: \B\ot \A^{\ot n}\to\B$,
where
$$\E^{(n+1)} = \left(\E^{(n)}\ot\id\right) \circ \E =
\left(\E\ot\id^{\ot n}\right) \circ \E^{(n)}
\quad,\deqno(iterE)$$
and $\E^{(1)} = \E$. Let $\rho$ be a density matrix on ${\cal M}_k$ and
identify $\rho$ with the state $B\in\B \mapsto \tr\rho B$. Given $\E$ and
$\rho$, we define a \cfc\ state $\om$ on $\A^-$ by:
$$\om(A) = \rho\left(\E^{(n)}(\idty_k\ot A)\right),\quad
A\in\A^{\bracks{-n,-1}}
\quad.\deqno(fcstateL)$$
Subscripts $d$ and $k$ of $\idty$ refer to $\A$ and $\B$ respectively. This
definition satisfies the compatibility condition $\om(A) = \om(\idty\ot A)$
because $\E$ is unity preserving. The Markovian or transfer matrix like
character of \cfc\ states can be put in evidence by expressing the
expectations of elementary tensors as:
$$\om(A_{-n}\ot A_{-n+1}\ot\cdots \ot A_{-1}) = \rho\left(\E_{A_{-1}}\circ
\E_{A_{-2}}\circ \cdots \circ\E_{A_{-n}}(\idty_k) \right)
\quad,\deqno(markov)$$
$A_{-i}\in\A$, $i=1,2,\ldots n$. The $\E_A$ in this formula are linear
transformations of $\B$ given by $\E_A(B) = \E(B\ot A)$, $B\in\B$. We will
mostly assume that the triple $(\B,\E,\rho)$ which generates $\om$ is
minimal in the sense that $\rho$ is a faithful state on $\B$ and that $\B$
is the smallest $\ast$-subalgebra of ${\cal M}_k$ containing $\idty_k$ and
invariant under the $\E_A$, $A\in\A$.

If $\rho$ satisfies the additional requirement:
$$\rho = \rho\circ\Eh
\quad,\deqno(rhoinv)$$
then $\om$ becomes a translation invariant state on the full chain
$\chain\A$ by putting
$$\om(A) = \rho\left(\E^{(n)}(\idty_k\ot A)\right),\quad
A\in\A^{\bracks{m,m+n}},\ m\in\Ir
\quad.\deqno(fcstateZ) $$
A distinctive role is played by the map $\Eh$ because its spectral
properties are directly connected to the ergodic properties of $\om$. In
\cite{FCS,FCP}, it was proven that a \cfc\ state $\om$ is ergodic iff
there exists a minimal generating triple $(\B,\E,\rho)$ for $\om$ such that
the eigenvector $\idty_k$ of $\Eh$ is non-degenerate. $\om$ is
exponentially clustering iff there is a minimal generating triple with
trivial peripheral spectrum, meaning that $\idty_k$ is the only eigenvector
of $\Eh$ with an eigenvalue of modulus 1. For general \cfc\ states, it is
not known whether minimal generating triples are unique, up to unitary
equivalence. Much more can be said if $\om$ is \pg, that is, if there is a
generating triple $(\B,\E,\rho)$ for $\rho$ with $\E = \ad(V^*)$ where $V$
is an isometry from $\Cx^k$ to $\Cx^k\ot\Cx^d$. Minimal generating triples
are unique in this case and a \cfc\ state $\om$ is pure iff it is \pg\ and
exponentially clustering. In this case $\B$ will automatically coincide
with the full ${\cal M}_k$. Furthermore, pure \cfc\ states arise as the
unique ground states of translation invariant, finite range interactions.

In order to make this connection more explicit we introduce the
iterates $V^{(n)}$ of $V$. $V^{(n)}$ is an isometry from $\Cx^k$
into $\Cx^k\ot(\Cx^d)^{\ot n}$, recursively defined by:
$$V^{(n+1)} = (V\ot(\idty_d)^{\ot n})\, V^{(n)} =
(V^{(n)}\ot\idty_d)\, V
\deqno(iterV)$$
and $V^{(1)}=V$. The $\E^{(n)}$ are now expressed as $\E^{(n)} =
\ad\left(V^{(n)*}\right)$. The reduced $n$-site density matrices
$\rho^{\bracks{1,n}}$ of $\om$ can easily be computed. For
$A\in\A^{\bracks{1,n}}$:
$$\eqalign{
\om(A)
&=\rho\left(V^{(n)*} \idty_k\ot A V^{(n)}\right) \cr
&=\tr \rho\left(V^{(n)*} \idty_k\ot A V^{(n)}\right) \cr
&=\tr \left(V^{(n)}\,\rho\,V^{(n)*}\, \idty_k\ot A\right)
\quad.}$$
Therefore
$$\rho^{\bracks{1,n}} = \tr_{\Cx^k} V^{(n)}\,\rho\,V^{(n)*}
\quad.$$

Let $\set{e_1,\ldots e_k}$ be an orthonormal basis of $\Cx^k$. It is clear
from the computation of above of that, for $0<n$, the reduced density
matrix $\rho^{\bracks{1,n}}$ will live on the subspace $\G_n$ of
$(\Cx^d)^{\otimes n}$ spanned by the vectors $\set{\phi_{ij}\stt
i,j=1,\ldots k}$, where:
$$V^{(n)}e_j = \sum_{i=1}^k e_i\ot\phi_{ij}
\quad.$$
Therefore $\rho^{\bracks{1,n}}$ is supported by a subspace $\G_n$ of
$(\Cx^d)^{\ot n}$ of dimension at most $k^2$, independently of $n$. It can
be shown that for $n$ large enough $\dim(\G_n)$ will eventually reach the
value $k^2$. Let $r$ be the smallest integer such that $\dim(\G_r)= k^2$.
If we choose as interaction $h\in({\cal M}_d)^{\ot(r+1)}$ the projection
operator in $(\Cx^d)^{\ot(r+1)}$ on the orthogonal complement of $\G_{r+1}$
then $\om(\tau^j(h))=0$ for all $j\in\Ir$. It is therefore a ground state
of $H=\sum_{j\in\Ir}\tau^j(h)$ in a very strong sense as it minimizes even
locally the energy. Moreover, it was shown in \cite{FCS} that $\om$ is
uniquely determined by the conditions $\om(\tau^j(h))=0$, $j\in\Ir$. This
means that $\om$ is ``locally'' exposed by the translates of $h$. The
interaction $h$ associated to the pure \cfc\ state $\om$ is often called a
VBS interaction.

We will now consider the construction of \cfc\ states and of VBS
interactions and ground states which are invariant under the action of a
quantum group \qg. Suppose that we are given unitary representations $v$
and $w$ of \qg\ on $\Cx^d$ and $\Cx^k$, implementing extended
actions $\ad(v)$ and $\ad(w)$ on $\A={\cal M}_d$ and on
$\B\subset{\cal M}_k$. A unity preserving \cp\ map $\E:\B\ot\A\to\B$
is {\it covariant\/} if:
$$\ad(w) \circ (\E\ot\id_\C) = (\E\ot\id_\C) \circ \ad(w\tee v)
\quad.\deqno(Ecov)$$
On the level of the iterates $\E^{(n)}$ of $\E$ covariance becomes:
$$\ad(w) \circ (\E^{(n)}\ot\id_\C) = (\E^{(n)}\ot\id_\C) \circ
\ad(w\tee v\tee\cdots \tee v)
\quad.\deqno(Encov)$$
It is instructive to write out the case $n=2$:
$$\eqalign{
\ad(w) \circ (\E^{(2)}\ot\id_\C)
&=\ad(w) \circ (\E\ot\id_\C) \circ (\E\ot\id_\A\ot\id_\C) \cr
&=(\E\ot\id_\C) \circ \ad(w\tee v) \circ (\E\ot\id_\A\ot\id_\C) \cr
&=(\E\ot\id_\C) \circ (\ad(w)\ot\id_\A) \circ (\id_\B\ot\ad(v)) \cr
      &\hskip 28pt \circ (\E\ot\id_\A\ot\id_\C) \cr
&=(\E\ot\id_\C) \circ (\ad(w)\ot\id_\A) \circ (\E\ot\id_\A\ot\id_\C) \cr
      &\hskip 28pt \circ (\id_\B\ot\id_\A\ot\ad(v)) \cr
&=(\E\ot\id_\C) \circ (\E\ot\id_\A\ot\id_\C)
      \circ(\ad(w)\ot\id_\A\ot\id_\A)               \cr
      &\hskip 28pt\circ (\id_\B\ot\ad(v)\ot\id_\A)
\circ (\id_\B\ot\id_\A\ot\alpha) \cr
&=(\E^{(2)}\ot\id_\C) \circ \ad(w\tee v\tee v)
\quad.}$$

The following Proposition shows how shift and quantum group invariance can
hold simultaneously for \cfc\ states restricted to a half chain, at the
cost of introducing an extra tensor factor, however.

\def\wo{\widetilde{\om}}
\iproclaim/fcs/ Proposition.
Let $(\B,\E,\rho)$ be a minimal triple generating the \cfc\ state $\om$
such that $\rho = \rho \circ \Eh$ and suppose that the eigenvalue 1 of
$\Eh$ is non-degenerate. Let $v$ and $w$ be unitary representations of a
quantum group \qg\ defining actions $\alpha_v$ and $\alpha_w$ on $\A$ and
$\B$ respectively and suppose that $\E$ is covariant.
\item{(1)}
  $\wo(X) = \rho\left(\E^{(n)}(X)\right)$, $X\in\B\ot \A^{\bracks{0,n-1}}$
  defines a state of $\B\ot\A^\Nl$ and $\wo$ coincides with $\om$ on
  $\idty_\B\ot\A^\Nl$.
\item{(2)}
  $\wo$ is invariant under the action $(\ad(w)\ot\id_{\A^\Nl}) \circ
  \alpha^\Nl_v$ on $\B\ot\A^\Nl$.
\eproclaim
\vskip -1truecm
\proof:
Let $X\in\B\ot\A^{\bracks{0,n-1}}$. We then compute:
$$\eqalign{
\rho\left(\E^{(n+1)}(X\ot\idty_d)\right)
&= \rho\left(\E\left(\E^{(n)}(X)\ot\idty_d\right)\right) \cr
&= \rho\left(\E^{(n)}(X)\right)
\quad.}$$
This is precisely the compatibility condition we need for $\wo$. Positivity
and normalization of $\wo$ are immediate consequences of the positivity and
normalization of $\rho$ and the $\E^{(n)}$. By construction $\wo$ extends
the restriction of $\om$ to $\A^\Nl$.

\noindent
We first show that for all $B\in\B$
$$(\rho\ot\id_\C)(\alpha_w(B)) = \rho(B) \idty_\C \quad.$$
Consider on $\B$ the functional
$$B \mapsto \rho\ot\sigma (\alpha_w(B)) \quad,$$
where $\sigma$ is an arbitrary continuous functional on $\C$.
Using the covariance of $\E$ and the invariance of $\rho$ under
$\Eh$ we compute:
$$\eqalign{
\rho\ot\sigma\bigl( \alpha_w(\E(B\ot\idty_d)) \bigr)
&=\rho\ot\sigma\bigl( \ad(w)\bigl( (\E\ot\id_\C) (B\ot\idty_d\ot\idty_\C)
\bigr) \bigr) \cr
&=\rho\ot\sigma\bigl( (\E\ot\id_\C)\bigl(\ad(w\tee v)
(B\ot\idty_d\ot\idty_\C)\bigr) \bigr) \cr
&=\rho\ot\sigma\bigl( (\E\ot\id_\C) (\alpha_w(B)\ot\idty_d)
\bigr) \cr
&=\rho\ot\sigma(\alpha_w(B))
\quad.}$$
By assumption, the eigenvalue $1$ of $\Eh$, and therefore also of its
dual, is non-degenerate. This implies that for all $\sigma\in\C^*$
$$\rho\ot\sigma (\alpha_w(B)) = \rho(B) \sigma(\idty_\C)
\quad.$$
Therefore
$$(\rho\ot\id_\C)(\alpha_w(B)) = \rho(B) \idty_\C $$
for $B\in\B$. The invariance of $\wo$ under $(\ad(w)\ot\idty_{\A^\Nl})
\circ \alpha_v^\Nl$ can now be checked. Let $X\in\B\ot\A^{\bracks{0,n-1}}$,
$n=1,2,\ldots$.
$$\eqalign{
\wo\ot\id_\C \bigl((\ad(w)\ot\id_{\A^\Nl}) (\alpha_v^\Nl(X)) \bigr)
\hskip-90pt&\hskip90pt
 =\wo\ot\id_\C \bigl(\ad(w\tee v\tee\cdots\tee v)
      (X\ot\idty_\C)\bigr) \cr
&=\rho\ot\id_\C \bigl( (\E^{(n)}\ot\id_\C)\bigl(
\ad(w\tee v\tee\cdots\tee v)
     (X\ot\idty_\C)\bigr)  \bigr) \cr
&=\rho\ot\id_\C \bigl(\ad(w)\bigl(\E^{(n)}(X)\ot\idty_\C\bigr)\bigr) \cr
&=\rho\ot\id_\C \bigl(\alpha_w\bigl(\E^{(n)}(X)\bigr)\bigr) \cr
&=\rho\bigl(\E^{(n)}(X)\bigr)\, \idty_\C \cr
&=\wo(X)\, \idty_\C
\quad.}$$
\QED

The density matrix $\rho$ used in the construction of the \cfc\ state
$\om$ in \Prp/fcs/ satisfies $\rho=\rho\circ\Eh$, which is needed in
order to insure the translation invariance of the state. It is
straightforward to check that there is another choice for the density
matrix, namely a density matrix $\rho^\prime$ that is invariant under
$\ad(w)$, which leads to an $\alpha^\Nl_v$ invariant state on
$\A^\Nl$. For quantum groups (that are not groups) one should not
expect these two requirements, invariance under $\ad(w)$ and
$\rho=\rho\circ\Eh$, to be compatible.  In the case of the
spin $S$ representation of \Sn2, $\rho$ and
$\rho^\prime$ are both unique and coincide only for $\q=1$:
$$ \rho={1\over [2S+1]_\q}\q^{-2\j_z} ,\quad \rho^\prime ={1\over
2S+1}\idty
$$
We refer to the Appendix for the notations and the calculation.
The conclusion of \Thm/qloc/, that $\alpha_v$ and translations are
incompatible properties, is then not so surprising.

One should note that, though $(\ad(w)\ot\id_{\A^\Nl}) \circ \alpha_v^\Nl$
is not a proper action on $\A^\Nl$, still, by \Thm/2pts/(2), if
$A\in\A^\Nl$ is invariant under $\alpha_v^\Nl$, $A$ is also invariant under
$(\ad(w)\ot\id_{\A^\Nl}) \circ \alpha_v^\Nl$. We now consider ground
states of spin chains corresponding to $\ad(v)$-invariant VBS interactions.
Suppose that we have two unitary representations $v$ and $w$ of \qg\ on
$\Cx^d$ and $\Cx^k$ respectively and an isometry $V: \Cx^k\to\Cx^k\ot\Cx^d$
intertwining $w$ and $v\tee w$, \ie:
$$(V\ot\idty_\C)\,w = (w\tee v)\,(V\ot\idty_\C)
\quad.\deqno()$$
The intertwining property on the level of the $V^{(n)}$ becomes:
$$(V^{(n)}\ot\idty_\C)\,w = (w\tee v^{\tee n})\,(V^{(n)}\ot\idty_\C)
\quad.\deqno()$$
Let $\rho$ be a density matrix on ${\cal M}_k$ such that
$$\rho(B) = \rho(V^*(B\ot\idty_d)V), \quad B\in{\cal M}_k
\quad.\deqno()$$
Generically, $\rho$ is uniquely determined by this condition and, putting
$\E=\ad(V^*)$, $\Eh$ has trivial peripheral spectrum. The \cfc\ state
generated by $({\cal M}_k,\E,\rho)$ is then a pure, translation-invariant
state on the chain $\chain\A$. Let, for $n=1,2,\ldots$, $\G_n$ be the
subspaces of $(\Cx^d)^{\ot n}$ introduced at the beginning of this section.
Recall that $\G_n$ is the supporting subspace of the reduced $n$-site
density matrix of $\om$.

\iproclaim/Gn/ Proposition.
Let $V:\Cx^k\to\Cx^k\ot\Cx^d$ be an isometry, intertwining the unitary
representations $v$ and $w$ of the quantum group \qg\ on $\Cx^d$ and
$\Cx^k$ respectively:
$$(V\ot\idty_\C)\,w = (w\tee v)\,(V\ot\idty_\C)
\quad.\deqno()$$
The orthogonal projection in $({\cal M}_d)^{\ot n}$ on the subspace
$\G_n$  of $(\Cx^d)^{\ot n}$ commutes with $v^{\tee n}$,
$n=1,2,\ldots$.
\eproclaim

\proof:
Let $\set{e_1,e_2,\ldots e_k}$ be an orthonormal basis for $\Cx^k$. The
subspace $\G_n$ of $(\Cx^d)^{\ot n}$ is generated by the vectors
$\set{\phi_{ij}\stt i,j=1,2,\ldots k}$ with
$$V^{(n)}e_j = \sum_{i=1}^k e_i\ot\phi_{ij}
\quad.$$
We can, without loss of generality, assume that $\C$ is a (norm-closed)
$\ast$-subalgebra of the bounded linear operators on some Hilbert space
$\H$. Let $\chi\in\H$. We have to show that, for $i,j=1,2\ldots k$,
$v^{\tee n} \phi_{ij}\ot\chi$ belongs to $\G_n\ot\H$ or, equivalently,
that for $j=1,2,\ldots k$
$$(\idty_k\ot v^{\tee n})\, (V^{(n)}\ot\idty_\C)\, e_j\ot \chi$$
is an element of $\G_n\ot\Cx^k\ot\H$.
$$\eqalign{
(\idty_k\ot v^{\tee n})\, (V^{(n)}\ot\idty_\C)\, e_j\ot \chi
&=(w^*\ot(\idty_d)^{\ot n})\, (w\tee v^{\tee n})\,
(V^{(n)}\ot\idty_\C)\,
e_j\ot \chi \cr
&=(w^*\ot(\idty_d)^{\ot n})\, (V^{(n)}\ot\idty_\C)\, w\, e_j\ot \chi
\quad.}$$
This proves the statement as $w^*$ acts only in a non-trivial way on
$\Cx^k\ot\H$.
\QED

In particular \Prp/Gn/ shows that the VBS interaction corresponding to
a pure \cfc\ state, generated by an isometric intertwiner of unitary
quantum group representations, is \qg-invariant. We conclude this
section with a discussion of what can be considered to be the simplest
possible example of this structure.


\noindent
{\it Example:} \nl
Consider the irreducible representations of \Sn2\ on $\Cx^2$
and $\Cx^3$. This leads to the $q$-deformed AKLT-model as considered
in~\cite{BY,Zitt} (in the present paper, however, the parameter is called
$\nu$ instead of $q$). Instead of using the Woronowicz description,
as in the example of section~2, we will turn to the Drinfel'd approach
that is much more effective for computations. The connection between
both approaches is sketched in the Appendix.

Denote by $[a]_\q$ the $\q$-numbers:
$[a]_\q=(\q^a-\q^{-a})/(\q-\q^{-1})$.  The commutation relations
between the ``Lie-algebra generators'' of \Sn2 are:
$$
[\j_z,\j_\pm]=
\pm\j_\pm \midbox{and} [\j_+,\j_-]= [2\j_z]_\q.
$$
Product
representations are constructed according to the rule
$$\eqalign{ L_z
&= \j_x\ot\idty+ \idty\ot K_z \cr L_\pm&= \q^{\j_z}\ot K_\pm+ \j_\pm\ot
\q^{-K_z}.  }
$$
There is a quadratic Casimir operator $C$ given by
$$C=
[\j_z+\half]_\q^2- [\half]_\q^2+ \j_-\,\j_+.
$$
The irreducible
representations of \Sn2 are completely similar to those of \SU2. There
is, for each $j\in\half\Nl$, a unique $(2j+1)$-dimensional
representation labelled by the eigenvalue $[j]_\q\,[j+1]_\q$ of $C$.
The explicit forms of the spin~$\half$ and spin~1 representations are:
$$
\j_z= \pmatrix{\half&0\cr0&-\half} \qquad \j_+= \pmatrix{0&1\cr0&0}
\qquad \j_-= \pmatrix{0&0\cr1&0}
$$
and
$$
\j_z=
\pmatrix{1&0&0\cr0&0&0\cr0&0&-1} \quad \j_+=
\pmatrix{0&\sqrt{[2]_\q}&0\cr0&0&\sqrt{[2]_\q}\cr0&0&0} \quad \j_-=
\pmatrix{0&0&0\cr\sqrt{[2]_\q}&0&0\cr0&\sqrt{[2]_\q}&0\cr}.  $$
Denoting by $\{\ket\half>,\ket-\half>\}$ and
$\{\ket1>,\ket0>,\ket-1>\}$ the canonical bases of $\Cx^2$ and $\Cx^3$,
the unique intertwiner $V$ between the spin~$\half$ representation and
the product of the spin~$\half$ and the spin~1 representation is easily
computed:
$$\eqalign{ V\,\ket\half> &=
{1\over\sqrt{[3]_\q}}\Bigl(\q^{-1}\,\ket\half,0>-
\q^{\half}\,\sqrt{[2]_\q}\,\ket-\half,1>\Bigr)\cr V\,\ket-\half> &=
{1\over\sqrt{[3]_\q}}\Bigl(\q^{-\half}\,\sqrt{[2]_\q}\,\ket\half,-1>-
\q\,\ket-\half,0>\Bigr).
}$$
The $2\times2$ density matrix $\rho$,
singled out by the invariance condition \eq(rhoinv), is
$$
\rho= {1\over[2]_\q} \pmatrix{\q^{-1}&0\cr0&\q},
$$
and the spectrum of $\Eh$
consist of 1 and $-\q^2/[3]_\q$, with degeneracy~3. The eigenvectors
are
$$
\idty,\quad \j_z \q^{2\j_z},\quad \j_+, \hbox{ and } \j_-,\quad .
$$
Finally, the \cfc\ state constructed in this way, is the unique,
shift-invariant ground state of the $\q$-invariant, nearest-neighbour,
VBS-Hamiltonian on the spin~1 chain, determined by the interaction
$h=C^2- [2]_\q\,C$. Here, the operator $C$ is the Casimir operator
in the tensor product of the spin~1 representation with itself.
It is, up to a normalization factor, the orthogonal projection onto the
spin~2 subrepresentation.
For $\q=1$ (and up to a multiplicative and additive
constant) $h$ reduces to the  well-known spin~1 AKLT-interaction
$3 \overline{\j_1}\cdot\overline{\j_2}+
(\overline{\j_1}\cdot\overline{\j_2})^2$, where $\overline{\j}$ denotes
the three Cartesian components of the spin~1 generators of \SU2\
\cite{AKLT}.

\vfill\eject
\bgsection A. Drinfel'd approach to $\SnU2$

The purpose of this Appendix is to set up the dual approach to
$\SnU2$, which is much more efficient in computations than the
Woronowicz version. Nobody in his right mind would do computations
concerning representations of $\rm{SU}(2)$ using the explicit form
of the representing unitaries as polynomials in matrix elements of
${\rm SU}(2)$. Yet this is what the Woronowicz approach requires.
Here we provide the associated Lie algebraic version of $\SnU2$, \ie
the corresponding object in the Drinfel'd approach. We present this
as a purely computational tool, and leave it to the reader to
construct the analogues of the results in the paper in this
language.

Throughout, we consider $\C_0$, the algebra of polynomials in the
generators $\alpha,\gamma$, and their adjoint and {\it not} the C*-algebra
of $\SnU2$. Likewise, tensor products are algebraic tensor products, and
the dual $\C_0^*$ is the algebraic dual. We make $\C_0^*$ into a Hopf
algebra with the operations
$$\eqalign{
    \xi\cdot\eta(a)         &= (\xi\otimes\eta)\circ\copr(a)  \cr
    \copr(\xi)(a\otimes b)  &= \xi\circ\multip(a\otimes b)
                             = \xi(ab)    \cr
    \idty(a)                &= \counit(a)
\quad.\cr}\deqno(dualHopf)$$
In the classical case there are two important kinds of linear
functionals on $\C(G)$: evaluations at group elements, and
directional derivatives at the identity. The latter make up the Lie
algebra, and, since $\C_0^*$ is an algebra, this space is to be
considered as the quantization of the universal enveloping algebra
of the Lie algebra of $\SnU2$, or the quantum group $\SnU2$ in the
sense of Drinfel'd.

We consider three special functionals $\j_z,\j_+,\j_-\in\C_0^*$, which
satisfy the relations
$$\eqalign{
      \copr(\j_z)&=\j_z\otimes\idty +\idty\otimes\j_z \cr
      \copr(\j_\pm)&=\j_\pm\otimes\q^{-\j_z} +\q^{\j_z}\otimes\j\pm
\quad.\cr}\deqno(CoprX)$$
Here the exponential is to be computed using the product in $\C_0^*$,
with the constant term given by the counit.
Using \eq(CoprX) we can compute these functionals on any
polynomials, once they are known on the generators. The following
table gives the necessary initial values.
$$ \matrix{
\bf \j(A) & A=\idty & A=\alpha &A=\alpha^* & A=\gamma& A=\gamma^*\cr
\j{}=\j_z &  0      & {1\over2}& -{1\over2}&   0    &   0        \cr
\j{}=\j_+ &  0      &     0    &   0       &   0     &-{1\over\q}\cr
\j{}=\j_- &  0      &     0    &   0       &   1     &  0        \cr
}\deqno(Xgen) $$
For $\j_z(A),\j_+(A)$, and $\j_-(A)$ to be well-defined on longer
products, we must guarantee that the value obtained using \eq(CoprX)
and \eq(Xgen) does not change if we transform $A$ by any of the
relations \eq(Snu2) of $\SnU2$ (including the relation
$\alpha\gamma=\q\gamma\alpha$). It suffices to show that if $A=0$ is
any of these relations, we get  $\j{}(A)=0$, and this is readily
verified. In particular, this fixes the relation between the
Woronowicz deformation parameter $\q$, and the parameter appearing
in \eq(CoprX).

The three functionals $\j\cdot$ are easy to compute directly on any
monomial. Let $m,m'$ denote monomials in $\alpha$ and $\alpha^*$,
and let $\abs m$ denote the grade of $m$ with respect to $\alpha$,
\ie the number of factors $\alpha$ minus the number of factors
$\alpha^*$. $\gamma^\sharp$ stands for either $\gamma$ or
$\gamma^*$, and $A\in\C_0$ is arbitrary. Then
$$\eqalign{
     \j_z(m\gamma^\sharp m')
          &=\q^{\pm\j_z}(m\gamma^\sharp m')=0 \cr
 \j_\pm(m\gamma^\sharp m')
          &=\q^{\j_z}(m)\ \j_\pm(\gamma^\sharp) \
                   \q^{-\j_z}(m')    \cr
     \j_z(m)      &= \abs m/2 \cr
 \q^{\pm\j_z}(m) &= \q^{\pm\abs m/2} \cr
    \j_\pm(m)      &=0              \cr
     \j_z(A^*)    &= -\Bar{\j_z(A)} \cr
     \j_+(A^*)    &= -{1\over\q} \Bar{\j_-(A)}
\quad.\cr}$$
Using the definition of the product in $\C_0^*$ in terms of the
coproduct of $\C$, we find the commutation relations
$$\eqalign{
   \bracks{\j_z,\j_\pm}&=\pm\j_\pm \cr
       \j_\pm\q^{\j_z}&=\q^{(\j_z \mp1)}\j_\pm  \cr
   \bracks{\j_+,\j_-}&=   {\q\over\q^2-1}\,
                       \Bigl( \q^{2\j_z} - \q^{-2\j_z}\Bigr)
\quad.\cr }\deqno(brackX)$$

Given a {\it unitary representation} $u\in\M d\C$, we can apply the
linear functionals $\xi\in\C_0^*$ to each matrix element, thus
obtaining a scalar matrix $\xi(u)$. Then the representation relation
\eq(repvij) becomes
$$   (\xi\cdot\eta)(u)=\xi(u)\eta(u)
\quad,$$
where on the left we have the product in $\C_0^*$, and on the right
the matrix product. In particular, the commutation relations
\eq(brackX) hold in any representation $u$. The unitarity of $u$
becomes a condition on the adjoints of the matrices $\j{}(u)$:
$$\eqalign{
    \j_z(u)^*&= \j_z(u)                     \cr
   \j_\pm(u)^*&=\j\mp(u)
\quad.}\deqno(uniX)$$

The condition of invariance of a state with respect to the action
$\alpha_u$ can be written directly in terms of the matrices
$\j{}(u)$ and the density matrix $\rho$. By applying $\j{}$ to the
equation $\sum_{i'j'}\rho_{i'j'}u_{j'j}(u_{i'i})^*=\rho_{ij}$, and
using the unitarity \eq(uniX), we get:
$$\eqalign{
                \j_z(u)\rho &= \rho\j_z(u)  \cr
   \j_+(u)\q^{2\j_z(u)}\rho &= \q^{2\j_z(u)}\rho\j_+(u)
\quad.}\deqno(invX)$$
In particular, $\rho=\q^{-2\j_z(u)}/[2S+1]_\q$ defines
an invariant state, and, for an irreducible representation this is
the only one.

\let\REF\doref
\Acknow
Most of this work was done during visits of M.F.\ to Osnabr\"uck. He would
like to take this opportunity to express his thanks for the warm
hospitality and most enjoyable cooperation. R.F.W.\ was partly supported by
a scholarship from the DFG (Bonn). We thank A. Van Daele for pointing out
to us the definition of compact quantum group and gratefully acknowledge
communications from Y. Watatani, and conversations with A.Yu.\ Alekseev, H.
Grosse, T. Matsui, K.-H. Rehren and P.~Vecserny\'es.

\refskip=12pt
\REF AF AFri   \Jref
     L. Accardi, A. Frigerio
    "Markovian Cocycles"
    \hfill\break
     Proc.R.Ir.Acad. @83A{(2)}(1983) 251--263

\REF AKLT AKLT      \Jref
     I. Affleck, T. Kennedy, E.H. Lieb, H. Tasaki
     "Valence bond ground states in isotropic quantum
     antiferromagnets"
     Commun.Math.Phys. @115(1988) 477--528

\REF ASW ASW \Gref
F.C. Alcaraz, S.R. Salinas, W.F. Wreszinski
"Anisotropic ferromagnetic quantum domains"
preprint

\REF BS Baaj \Jref
    S. Baaj, G. Skandalis
    "Unitaires multiplicatifs et dualit\'e pour les produits crois\'es
    de C*-alg\`ebres"
    Ann.Ec.Norm.Sup. @26(1993) 425

\REF Bab   BAB       \Jref
     H.M. Babujian
     "Exact solution of the isotropic Heisenberg chain with arbitrary spins:
     Thermodynamics of the model"
     Nuclear Phys.B @215(1983) 317--336

\REF BMNR BMNR \Jref
    M.T. Batchelor, L. Mezincescu, R. Nepomechie, V. Rittenberg
    "q-Deforma\-tions of the O(3)-symmetric spin-1 chain"
    J.Phys. @A23(1990) L141--L144

\REF BY BY \Gref
     M.T. Batchelor, C.M. Yung
     "q-Deformations of quantum spin chains with exact valence bond
      ground states"\hfill\break
      preprint archived in {\tt cond-mat@babbage.sissa.it \#9403080}

\REF BF Bernard \Jref
    D. Bernard, G. Felder
    "Quantum group symmetries in two-dimensional lattice quantum
    field theory"
    Nucl.Phys.B @365(1991) 98--120

\REF Ber Bernard2 \Jref
    D. Bernard
    "Quantum group symmetries and non-local currents in 2D QFT"
    Commun.Math.Phys. @142(1991) 99--138

\REF Bie Biedenharn \Gref
     L.C. Biedenharn
    "A q-Boson realization of the quantum group SU$_q$(2), and the
    theory of q-tensor operators"
    pp. 67-88
    \inPr H.-D. Doebner, J.-D. Hennig
    "Quantum groups"
    Springer Lect.Not.Phys. {\bf 370}, Berlin 1990

\REF BR BraRo      \Bref
    O. Bratteli, D.W. Robinson
     "Operator algebras and quantum statistical mechanics"
     2 volumes, Springer Verlag, Berlin, Heidelberg, New York
     1979 and 1981

\REF CE CEffros \Jref
    M.D. Choi, E.G. Effros
    "Nuclear C*-algebras and the approximation property"
    Ann.Math. @100(1978) 61--79

\REF Cu1 Cuntz  \Jref
    J. Cuntz
    "Simple C*-algebras generated by isometries"
    \hfill\break
    Commun.Math.Phys. @57(1977) 173--185

\REF Cu2 Cuntzact \Gref
    J. Cuntz
    "Regular actions of Hopf algebras on the C*-algebra generated by
    a Hilbert space"
    \inPr R. Herman, B. Tanbay
    "Operator algebras, mathematical physics, and low dimensional
    topology"
    A.K. Peters, Wellesley 1993

\REF DC Dasgupta \Jref
    N. Dasgupta, A.R. Chowdhury
    "Algebraic Bethe ansatz with boundary condition for
    SU$_{p,q}$(2) invariant spin chain"
    J.Phys.A @26(1993) 5427--5433

\REF DFJMN DFJMN \Jref
     B. Davies, O. Foda, M. Jimbo, T. Miwa, A. Nakayashiki
     "Diagonalization of the XXZ Hamiltonian by Vertex Operators"
     Commun.Math.Phys. @151(1993) 89--153

\REF DHR DHR \Jref
    S. Doplicher, R. Haag, J.E. Roberts
   "Fields, observables and gauge transformations, I"
    Commun.Math.Phys. @13(1969) 1--23
    \more; \hfill\break Part II in
    \Jn Commun.Math.Phys. @15(1969) 173--200

\REF DR1 Roberts \Jref
    S. Doplicher, J.E. Roberts
    "Endomorphisms of C*-algebras, cross products and duality
    for compact groups"
    Ann.Math. @130(1989) 75--119

\REF DR2 Roberts2 \Gref
    S. Doplicher, J.E. Roberts
    "C*-algebras and duality for compact groups: Why there is a
    compact group of internal symmetries in particle physics"
    pp. 489-498
    \inPr R. S\'en\'eor and M. Mebkhout
    "Proceedings of the International Conference on Mathematical
    Physics"
    World Scientific, Singapore 1986

\REF Dri Drinfeld \Gref
    V.G. Drinfel'd
    "Quantum groups"
    pp. 798-820 in vol.1 of the Proceedings of the Int.Congr.Math.
    Berkeley 1986,
    Academic Press 1987

\REF DW LMD \Jref
     N.G. Duffield, R.F. Werner
     "Local dynamics of mean-field quantum systems"
     Helv.Phys.Acta @65(1992) 1016--1054

\REF FNW1 FCS      \Jref
    M. Fannes, B. Nachtergaele, R.F. Werner
    "Finitely correlated states of quantum spin chains"
    Commun.Math.Phys. @144(1992) 443--490

\REF FNW2 FCP      \Jref
    M. Fannes, B. Nachtergaele, R.F. Werner
    "Finitely correlated pure states"
    J.Funct.Anal. @120(1994) 511--534

\REF FSV FSV  \Jref
   M. Fannes, H. Spohn, A. Verbeure
   "Equilibrium states for mean field models"
   J. Math. Phys. @21(1980) 355--358

\REF GW GW \Gref
C.-T.Gottstein and R.F. Werner
"Ground states of the infinite q-deformed Heisenberg ferromagnet"
archived in {\tt cond-mat@babbage.sissa.it \#9501123}

\REF GS Grosse \Gref
    H. Grosse, E. Raschhofer
    "Bethe-Ansatz solution of a modified SU(3)-XXZ model"
    pp. 385-392
    \inPr M. Fannes, C. Maes, A. Verbeure
    "On three levels; micro-, meso-, and macro-approaches in physics"
    Plenum Press, New York 1994

\REF Ha1 Haldane \Jref
      F.D.M. Haldane
      "Exact Jastrow-Gutzwiller resonating-valence-bond ground\hfill\break
      state of the spin-1/2 anti\-ferro\-magnetic
      Heisen\-berg chain with $1/r^2$ exchange"
      Phys.Rev.Lett. @60(1988) 635--638

\REF Ha2 Haldanereview \Gref
      F.D.M. Haldane
      "Physics of the ideal semion gas: spinons and quantum
      symmetries of the integrable Haldane-Shastry spin chain"
      to appear in the Proceedings of the 16th Taniguchi Symposium,
Kashikojima,
      Japan, edited by A. Okiji and N. Kawakami, Springer,
      Berlin-Heidelberg-New York,1994

\REF Jim Jimbo \Jref
    M. Jimbo
    "A q-Difference Analogue of $U(g)$ and the Yang-Baxter equation"
    Lett. Math. Phys. @10(1985)63-69

\REF JSW WICK \Gref
   P.E.T. J\o rgensen, L.M. Schmitt, R.F. Werner
   "Positive representations of general commutation relations
    allowing Wick ordering"
   Preprint Osnabr\"uck and Iowa, 1993
   archived in {\tt funct-an@babbage.sissa.it \#9312004}

\REF KSZ Zitt \Jref
    A. Kl\"umper, A. Schadschneider, J. Zittartz
    "Groundstate properties of a generalized VBS-model"
    Zeitschrift f\"ur Physik B  @87(1992) 281--287

\REF KNW Watatani \Jref
    Y. Konishi, M. Nagisa, Y. Watatani
    "Some remarks on actions of compact matrix quantum groups on
    C*-algebras"
    Pacific.J.Math. @153(1992) 119--128

\REF KS Kulish \Jref
    P.P. Kulish, E.K. Sklyanin
    "The general $U_q\lbrack sl(2)\rbrack$ invariant XXZ integrable
    quantum spin chain"
    J.Phys.A @24(1991) L435--L439

\REF LB Ma \Jref
    C.R. Lienert, P.H. Butler
    "Racah-Wigner algebra for q-deformed algebras"
    J.Phys.A @25(1992) 1223--1235

\REF MS Mack \Jref
    G. Mack, V. Schomerus
    "Conformal field algebras with quantum symmetry from the theory
    of super\-selec\-tion sec\-tors"
    Commun.Math.Phys. @134(1990) 139--196

\REF Maj Majid \Gref
    S. Majid
    "Braided matrix structure of the Sklyanin algebra and of the quantum
    Lorentz group"
    Commun.Math.Phys. @156(1993) 607--638

\REF Mat Mathematica \Bref
   \noinitial. Wolfram Research{,} Inc.
   "Mathematica 2.2"
    Wolfram Research, Inc., Champaign, Illinois 1992

\REF MMP Zagreb \Jref
    S. Meljanac, M. Milekovi\'c, S. Pallua
    "Deformed SU(2) Heisenberg chain"
    J.Phys. A @24(1991) 581--591

\REF MN MezNep \Jref
    L. Mezincescu, R.I. Nepomechie
    "Analytical Bethe Ansatz for quantum-algebra-invariant
    spin chains"
    Nucl.Phys. B @372(1992) 597--621

\REF RW MF \Jref
     G.A. Raggio, R.F. Werner
     "Quantum statistical mechanics of general mean field
         systems"
     Helv.Phys.Acta @62 (1989) 980--1003

\REF Rue Ruegg \Jref
   H. Ruegg
   "A simple derivation of the quantum Clebsch-Gordan coefficients
   for SU(2)$_q$"
   J.Math.Phys. @31(1991) 1085--1087

\REF Sha Shastri \Jref
    B.S. Shastri
    "Exact solution of an $S=1/2$ Heisen\-berg anti\-ferro\-magnetic chain
     with long-ranged interactions"
    Phys.Rev.Lett. @60(1988) 639--642

\REF SV SV \Jref
     K. Szlach\'anyi, P. Vecserny\'es
     "Quantum symmetry and braid group statistics in $G$-spin models"
     Commun.Math.Phys. @156(1993) 127--168

\REF Tak Takesaki      \Bref
    M. Takesaki "Theory of operator algebras I"
    Springer Verlag, Berlin, Heidelberg, New York 1979

\REF Vec Vecser \Gref
      P. Vecserny\'es
     "On the quantum symmetry of the chiral Ising model"
     Princeton U. preprint PUPT-1406, archived in
     {\tt hepth@xxx.lanl.gov \#9306118}

\REF Wo1 WoRims  \Jref
    S.L. Woronowicz
    "Twisted SU(2) group. An example of a non-commutative
    differential calculus"
    Publ.RIMS, Kyoto @23(1987) 117--181

\REF Wo2 Woronowicz \Jref
    S.L. Woronowicz
    "Compact matrix pseudogroups"
    \hfill\break
    Commun.Math.Phys. @111(1987) 613--665

\REF Wo3 WoronDC \Jref
    S.L. Woronowicz
    "Differential calculus on compact matrix pseudogroups (quantum
    groups)"
    Commun.Math.Phys. @122(1989) 125--170

\REF Wo4 WoronCG \Gref
    S.L. Woronowicz
    "Compact quantum groups"
    In preparation.

\bye